\documentclass[aps,prx,reprint,superscriptaddress,floatfix,showpacs,longbibliography,nofootinbib]{revtex4-1}
\usepackage{url}
\usepackage{graphicx,amssymb,amsfonts,amsmath}
\usepackage[caption=false]{subfig}
\def\Label#1{}
\usepackage{graphicx}
\usepackage{amssymb,amsmath,bm}
\usepackage{color}
\usepackage{mhequ}
\usepackage{times}
\usepackage{amsthm}
\usepackage{accents}
\usepackage{blkarray}

\def\fref#1{Fig.~\ref{#1}}
\let\phi=\varphi
\let\rho=\varrho

\def\eref#1{Eq(\ref{#1})}

\def\cref#1{Corollary~\ref{#1}}

\def\sref#1{Sect.~\ref{#1}}

\def\half{{\textstyle{\frac{1}{2}}}}
\def\ie{{\it i.e.}}
\def\eg{{\it e.g.}}
\def\HH{{\mathcal{H}}}

\newenvironment{myitem}
{\begin{itemize}
  \setlength{\itemsep}{1pt}
  \setlength{\parskip}{0pt}
  \setlength{\parsep}{0pt}}
{\end{itemize}}

\newenvironment{myenum}
{\begin{enumerate}
  \setlength{\itemsep}{1pt}
  \setlength{\parskip}{0pt}
  \setlength{\parsep}{0pt}}
{\end{enumerate}}

\def\emphx#1{\emph{#1}\index{#1}}
\begin{document}

\title{                                                                                                                  Physical model of the genotype-to-phenotype map of proteins}
\author{Tsvi Tlusty}
\affiliation{Center for Soft and Living Matter, Institute for Basic Science (IBS), Ulsan 44919, Korea}
\affiliation{Department of Physics, Ulsan National Institute of Science and Technology (UNIST), Ulsan 44919, Korea}
\affiliation{Simons Center for Systems Biology, Institute for Advanced Study, Princeton, NJ 08540, USA}
\author{Albert Libchaber}
\affiliation{The Rockefeller University, 1230 York Avenue, New York, NY 10021, USA }
\author{Jean-Pierre Eckmann}
\affiliation{D\'{e}partement de Physique Th\'{e}orique and Section de
Math\'{e}matiques, Universit\'{e} de Gen\`{e}ve, CH-1211, Geneva 4,
Switzerland}

\begin{abstract}
How DNA is mapped to functional proteins is a  basic question of living matter. We introduce and study a physical model of protein evolution which suggests a mechanical basis for this map. Many proteins rely on  large-scale motion to function. We therefore  treat
protein as learning amorphous matter that evolves towards such a  mechanical function: Genes are  binary sequences that encode the connectivity of the amino acid network that makes a protein.  The gene is evolved until the network forms a shear band across the protein, which allows for long-range, soft modes required for protein function. The evolution reduces the high-dimensional sequence space to a low-dimensional space of mechanical modes, in accord with the observed dimensional
reduction between genotype and phenotype of proteins.  Spectral
analysis of the space of $10^6$ solutions shows a strong
correspondence between localization around the shear band of both
mechanical modes and the sequence structure. Specifically, our model shows how mutations are correlated among amino acids
whose interactions determine the functional mode.

\end{abstract}


\maketitle 

\section{Introduction: proteins and the question of the genotype-to-phenotype map}
\label{sec:introduction}
DNA genes code for the three-dimensional configurations of amino acids that make functional proteins. This sequence-to-function map is hard to decrypt since it links the collective physical interactions inside
the protein to the corresponding evolutionary forces acting on the gene \cite{Koonin2002,Xia2004,Dill2012,Zeldovich2008,Liberles2012}. Furthermore, evolution has to select the tiny fraction of functional sequences in an enormous, high-dimensional space \cite{Povolotskaya2010,Keefe2001,Koehl2002}, which implies that protein is a non-generic, \emph{information-rich} matter, outside the scope of standard statistical methods. Therefore, although the structure and physical forces within a protein have been extensively studied, the fundamental question as to how a functional
protein originates from a linear DNA sequence is still open, in particular, how the functionality constrains the accessible DNA sequences. 

To  examine the geometry of the sequence-to-function map, we devise a mechanical model of proteins as amorphous learning matter. Rather than simulating concrete proteins, we construct a model which captures the hallmarks of the genotype-to-phenotype map. The model is simple enough to be efficiently simulated to gain statistics and insight into the geometry of the map. We base our model on the growing evidence that large-scale conformational changes \--- where big chunks of the protein move  with respect to each other \--- are central to function
\cite{Koshland1958,Henzler-Wildman2007,Savir2007,Schmeing2009,Savir2010a,Huse2002,Savir2013}. In particular, allosteric proteins  can be viewed as `mechanical transducers' that transmit regulatory signals between distant sites \cite{Perutz1970,Goodey2008,Lockless1999,Ferreon2013}. 

Dynamics is essential to protein function, but it is hard to measure and  simulate  due to the challenging spatial and temporal scales. Nevertheless, recent studies suggest  a physical picture of the functionally-relevant  conformational changes within the protein: Nanorheological 
measurements showed low-frequency viscoelastic flow within enzymes \cite{Qu2013}, with  mechanical stress affecting catalysis \cite{Joseph2014}. Computation of amino acid  displacement, by analysis of structural data, demonstrated  that the strain is localized  in 2D  bands across allosteric enzymes   \cite{Mitchell2016}. 
We therefore take as a target function to be evolved in our protein such a large-scale dynamical mode. Other important functional constraints, such as specific chemical interactions at binding sites,  are disregarded here because they are confined to a small fraction of the protein. We focus on this mechanical function  whose  large scale, collective nature leads to long-range correlation patterns in the gene.

Our model includes essential elements  of the genotype-to-phenotype map: the target mechanical mode is evolved by mutating the `gene' that determines the connectivity in the amino acid network. During the simulated `evolution', mutations eventually divide the protein into rigid and `floppy' domains, and this division enables large-scale motion in the protein \cite{Gerstein1994}. This provides a concrete map between sequence, configuration, and function of the protein. The  computational simplicity allows for a massive survey of the sequence universe, which reveals a strong signature of the protein's structure and function within correlation `ripples' that appear in the space of DNA sequences.

\begin{figure}[ht]
\includegraphics[width=\columnwidth]{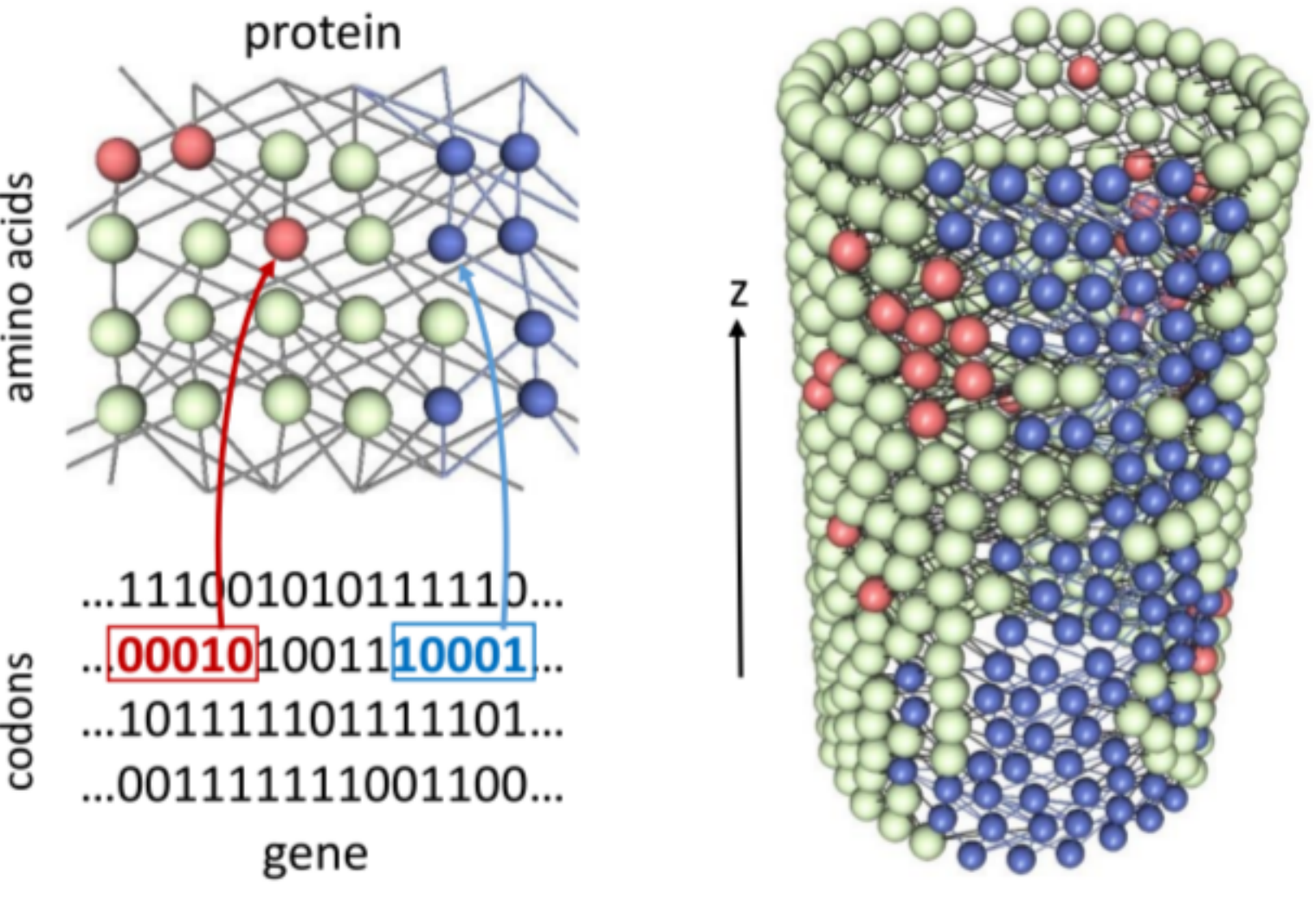}
\caption{
\textbf{The main features of the physical model}:\\
(left) The mapping from the binary gene to the connectivity  of the amino acid (AA)\ network that makes a functional protein. AAs are beads and links are bonds. The  
color of the AAs represents their rigidity state as determined  by the connectivity according to the algorithm of \sref{sec:rigidity}. \
Each AA can be in one of three states: rigid (gray) or fluid (\ie, non-rigid), which are divided between shearable (blue) and non-shearable (red). \\ 
(right) The AAs in the model protein are arranged in the shape of a cylinder, in this case with a
fluid channel (blue region). Such a configuration  can transduce a mechanical signal of shear or hinge motion along the fluid channel.\ }\label{fig:fig1} \end{figure}

\begin{figure}[ht]
\includegraphics[width=\columnwidth]{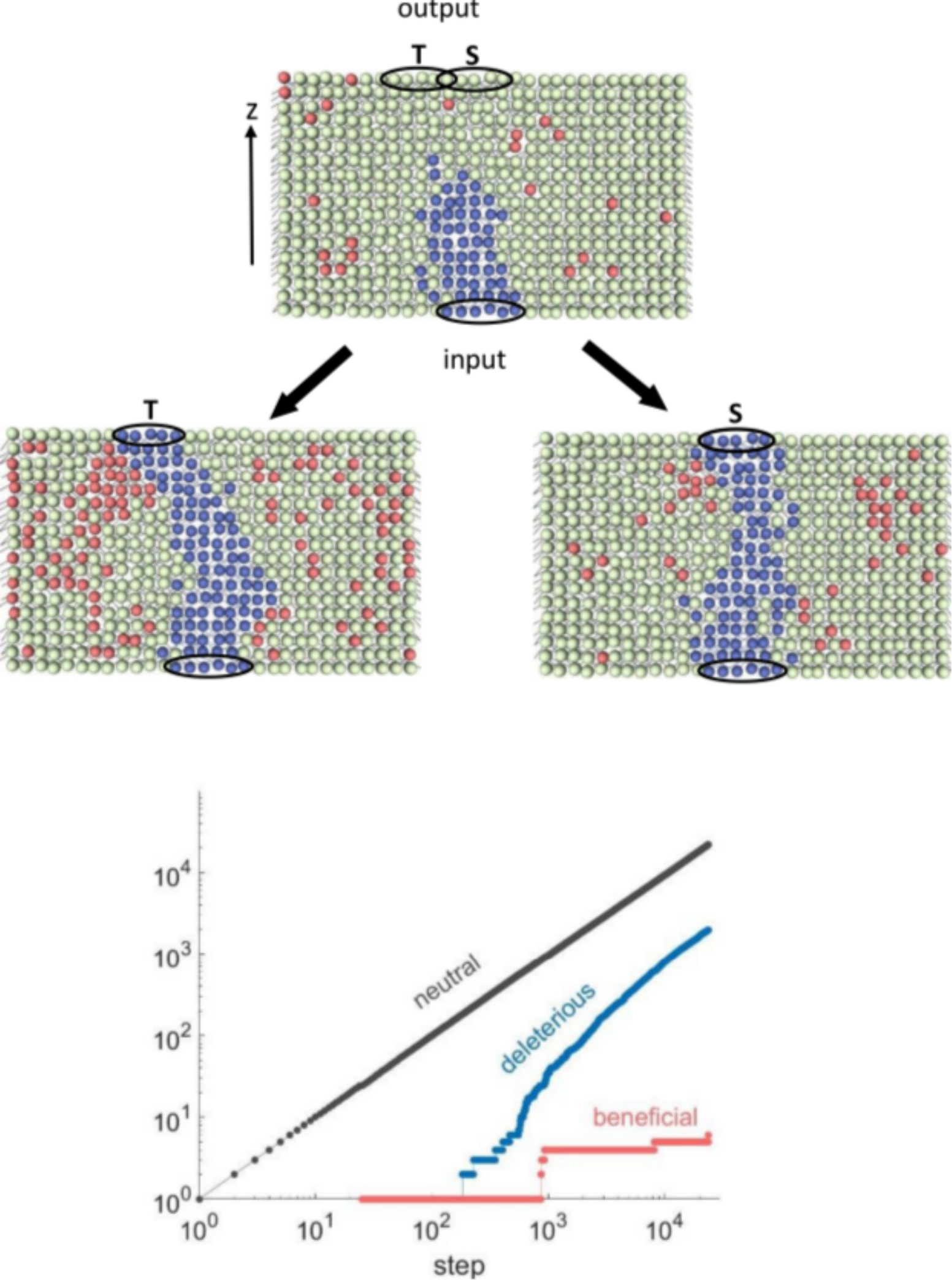}
\caption{
{\bf Evolution of mechanical function:}\\
(top) An initial configuration with a given input (black
ellipse at bottom) and a random sequence is required to evolve into a straight fluid channel (S) or a tilted one (T).\\
(bottom) Following the success of evolution. In
each generation, a randomly drawn bit (a letter in the 5-bit codon) is flipped, and this `point mutation' is changing one bond (similar to point mutations that change one base in a codon). A typical run is a sequence of mostly neutral steps, a fraction of deleterious ones, and rare beneficial steps. Note that the `fitness' of the configuration  is only measured at the top, not in the interior of the cylinder.
}\label{fig:fig1b} 
\end{figure}

\section{Model and  results}
\label{sec:model}

We give here a summary and interpretation of our results,   The appendix contains further details and explains choices we made in designing the model as close as
possible to real proteins.

\subsection{Mechanical model of protein evolution}
\label{sub:mechanical model}
Our model is based on two structures: a \emph{gene}, and a  \emph{protein}, which are coupled by the genotype-to-phenotype map.  
The coarse-grained protein is an aggregate of amino acids (AAs), modeled as beads, with short-range interactions  given  as bonds (\fref{fig:fig1}). A typical protein is made of several hundred AAs, and we take $N=540$. We layer the AAs on a cylinder, $18$ high  $30$ wide, similar to dimensions of globular proteins. The cylindrical configuration allows for fast calculation of the low energy modes, and thereby fast evolution of the protein.
Each AA may connect to the nearest five AAs in the layer below, so that we get $2^5 = 32$ effective AA species, which are encoded as 5-letter binary \emph{codons} 
\footnote{In our model, the AA species is determined by the bonds, while in real proteins the bonds are determined by the chemical nature and position of the AA (see also \sref{sub:aa}).}. 
These codons specify  the bonds in the protein  in a 2550-long \emph{sequence} of the \textit{gene}($5\times30
\times(18-1)$, because the lowest layer is connected only upwards).

To become functional, we want the protein to evolve to a configuration of AAs and bonds that can
transduce a mechanical signal from a prescribed input at the bottom of the cylinder to a prescribed output at its top 
\footnote{Note that in this simulation, we do not take as evolutionary criterion the mechanical signal itself, but require  that the protein forms a fluid channel with a prescribed configuration. We show that this configuration  \emph{facilitates} the sought-after mechanical shear motion in Sections \ref{sub:fluid} and \ref{subs}. (In \cite{Dutta2017} we
take the mechanical modes themselves as the
target function.)}.
The solution we search turns out to be a large-scale, low-energy deformation where one domain moves rigidly
with respect to another in a shear or hinge motion, which is facilitated by the presence of a fluidized, `floppy' channel separating the rigid domains 
\cite{Alexander1983,Phillips1985,Alexander1998}. 

These large-scale deformations are governed by the rigidity pattern of the configuration, which is determined by the connectivity of the AA network via a simple majority rule (\fref{fig:fig1}) which we detail in \sref{sec:rigidity}. The basic idea is that each AA can be either rigid or fluidized and that this  rigidity state  propagates upwards: Depending on the number of bonds and the state of other AAs in its immediate neighborhood, an AA will be rigidly connected, `shearable', \ie, loosely connected, or in a pocket of less connected AAs within a rigid neighborhood 
\footnote{The propagation of rigidity is effectively a ``double" percolation problem in which both fluid (blue) and rigid (gray) regions are continuous (see \sref{sec:rigidity}).}. 
As the sequence and hence the connections mutate, the model protein adapts to the desired input-output relation specified by the extremities of the separating fluid channel (\fref{fig:fig1}(right)).

The model is easy to simulate: We start from a random gene of $2550$ bits, and at each time step 
we flip a randomly drawn bit, thus adding or deleting a bond. In a zero-temperature Metropolis fashion, we keep only mutations which 
do not increase the distance from the target function, \ie, the number of errors between
the state in the top row and the prescribed outcome. Note that, following the logics of biological evolution, the `fitness' of the protein is only measured at its functional \emph{surface} (\eg, where a substrate binds to an enzyme) but not in its interior.

Typically, after $10^3$-$10^5$ mutations this
input-output problem is solved (\fref{fig:fig1b}). Although the functional sequences are extremely sparse among the $2^{2550}$ possible sequences, the small bias for getting closer to the target in configuration space directs the search rather quickly. Therefore, we could calculate as much as $10^6$ runs of the simulation which gave  $10^6$ independent solutions of the evolutionary task.

\begin{figure}\includegraphics[width=\columnwidth]{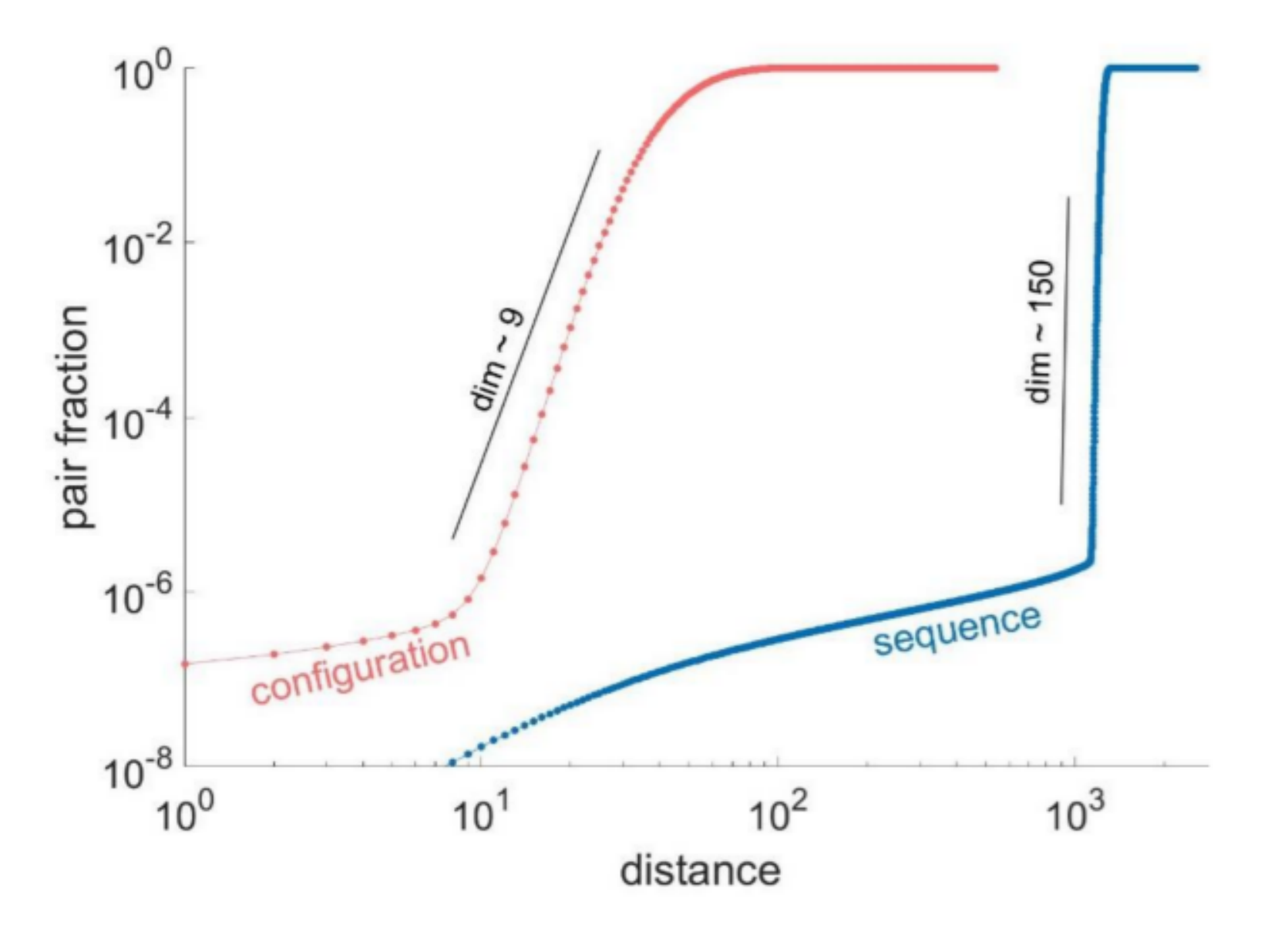}
\includegraphics[width=\columnwidth,angle=0]{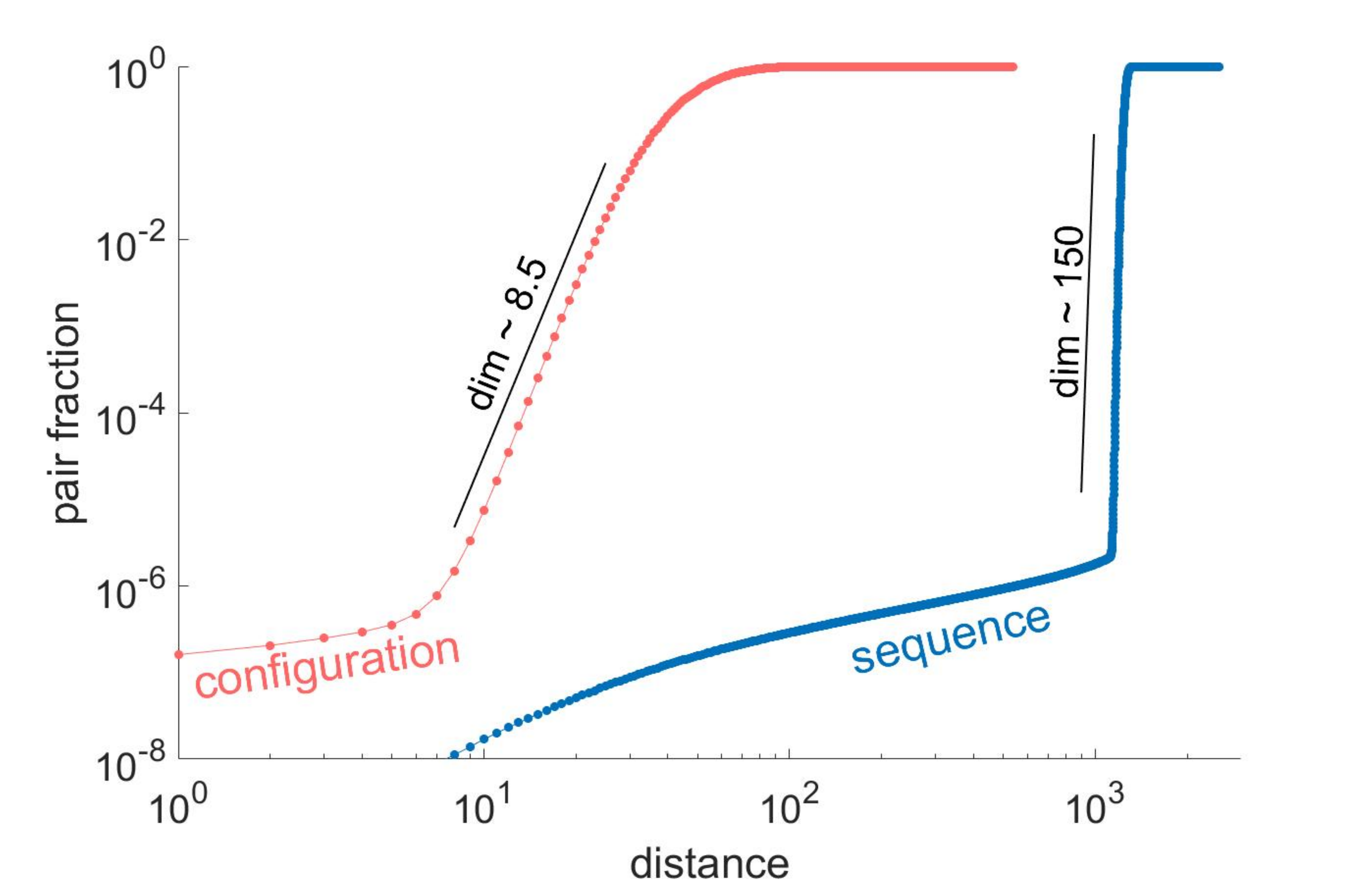}
\caption{
\textbf{Dimensional reduction of the genotype-to-phenotype map:}\\
Dimension measurement for the straight (S, top) and tilted (T, bottom) cases. $10^6$ independent functional configurations were found for the input-output problem. An estimate for the dimension of the solutions is the correlation length, the slope of the cumulative fraction of solution pairs as a function of distance. In
configuration space (red), the distance is the number of AAs (out of 540) with a different rigidity state. The estimated dimension from $10^{12}/2$ distances is about $9$ (black line) for problem S and $8.5$ in problem T. \\ The sequence space is a 2550-dimensional
hypercube with $32^{510}$ sequences. Most distances are close to the typical distance between two random sequences (2550/2 = 1275), indicating a high-dimensional solution space. An estimate for the dimension is $\sim150$ (black line) for both S\ and T\ problems. The similarity of the dimensions in both cases  suggests that these numbers are not specific to the problem. \\   
}\label{fig:fig2} 
\end{figure}

%

\begin{figure}\includegraphics[width=\columnwidth]{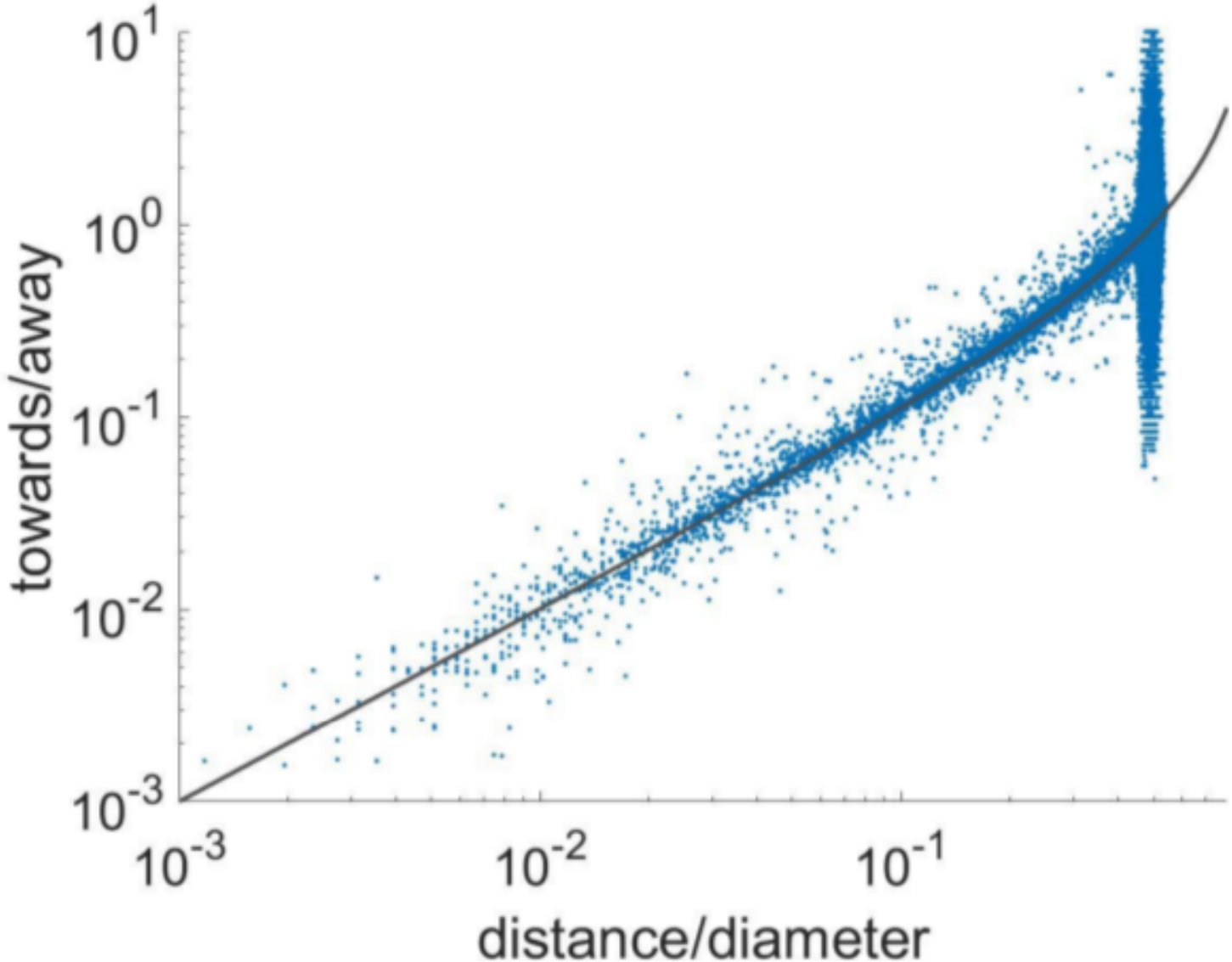}
\caption{
\textbf{Distribution of solutions in the sequence universe: }\\
A measure for the expansion in the functional sequence universe is the backward/forward ratio, the
fraction of point mutations that make two sequences closer vs.~the ones that increase the distance \cite{Povolotskaya2010}. The Hamming distances $D$ (normalized by the universe diameter $d_{\max}= 2550$ ) show that most sequences reach the edge of the universe, where no further expansion is possible. The black curve, $D/(1-D)$, is the backward/forward ratio from purely random mutations.
}\label{fig:fig2b} 
\end{figure}


\begin{figure}[h]\includegraphics[width=\columnwidth]{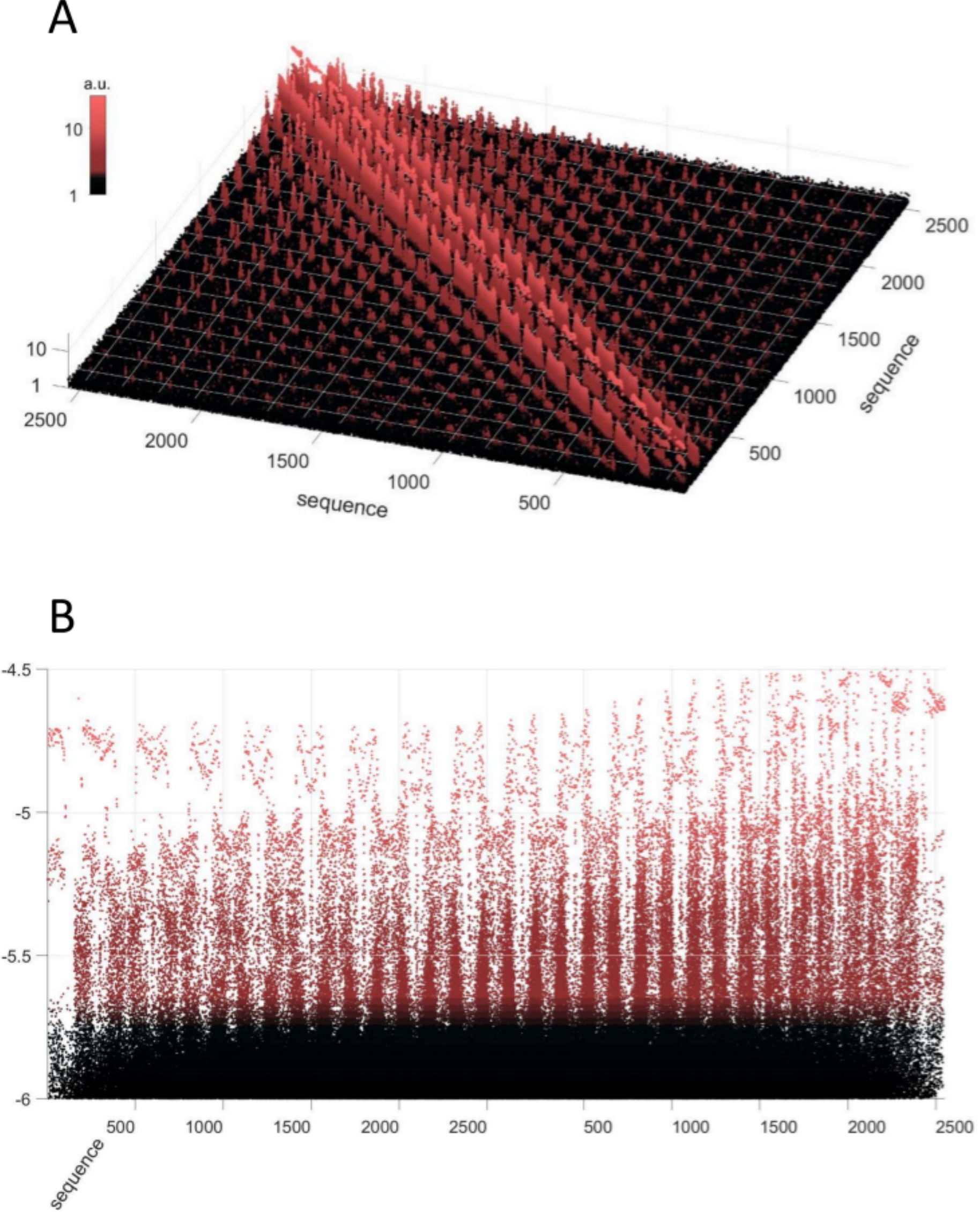}
\caption{
\textbf{Long-range genetic correlations:}\\
(top) The sequence correlation matrix across the $10^6$ examples shows long-range correlations among the bits (codons) at the rigid/fluid boundary, and short-range
correlations in the rigid domains.\\
(bottom) A cross section perpendicular to the diagonal  axis.}
\label{fig:fig2c}
\end{figure}

\subsection{Dimensional reduction in the phenotype-to-genotype map}
\label{sub:dimensional}
Thanks to the large number of simulations, we can
explore vast regions of the genetic universe. That the sampling is well-distributed  can be seen from
the typical inter-sequences distance, which is comparable with the universe diameter (\fref{fig:fig2b}).
This also indicates that the dimension of the solution set is high. Indeed, the observed dimension
of \emph{sequence} space, as estimated following  \cite{Procaccia1988,Eckmann1992}, is practically infinite ($\sim150$) 
\footnote{We lack sufficient data to determine such high dimensions precisely, and $150$ is a lower bound.}. This shows that the bonds are chosen basically at
random, although we only consider functional sequences.

On the other hand, very few among the $2^{540}$ \emph{configurations} are solutions, owing to the physical constraints of contiguous rigid and shearable domains. As a result, when mapped to the configuration space, the
solutions exhibit a dramatic reduction to a dimension of about 8-10 \cite{Grassberger1983}. This reduction between `genotype' (sequence) and `phenotype' (configuration, function) \cite{Savir2010,Shoval2012} is the outcome of physical constraints on the mechanical transduction problem. In the nearly random background of sequence space, these constraints are also manifested in long-range
correlations among AAs on the boundary of the shearable region (\fref{fig:fig2c} and \sref{sec:correlation}).  

\begin{figure}\includegraphics[width=\columnwidth]{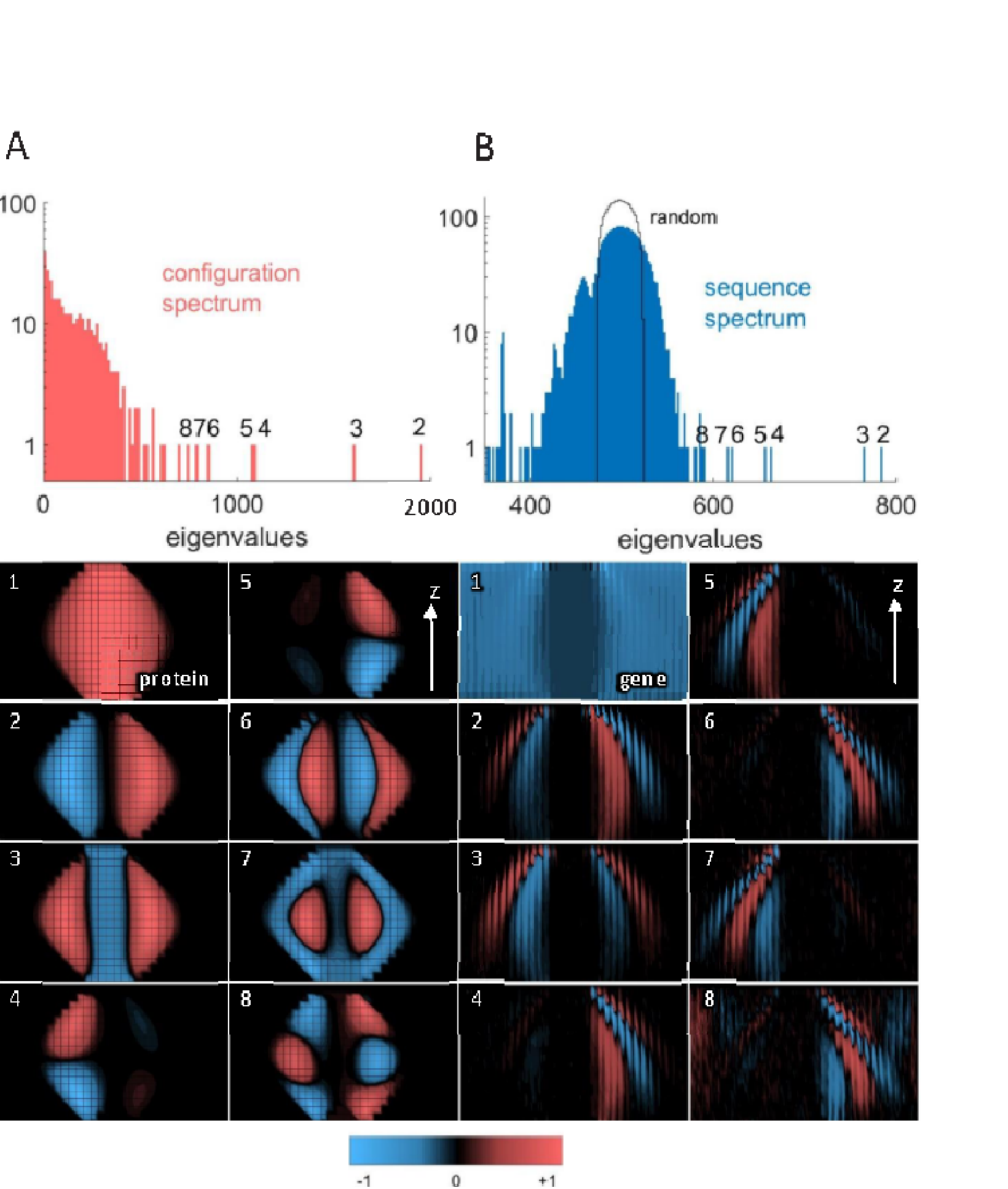}
\caption{\label{fig:fig3}\textbf{Correspondence of modes in sequence and configuration
spaces:}\\
 We produced the spectra by singular value decomposition of the $10^6$ solutions of problem S. The corresponding spectra for the T   case are shown in \fref{fig:tspec}.\\
(A) Top: the spectrum in configuration space exhibits about 8-10 eigenvalues outside the continuum (large $1^{\text{st}}$ eigenvalue not shown).\\  Bottom: the corresponding eigenvectors describe the basic modes of the fluid channel, such as side-to-side shift ($2^{\text{nd}}$) or expansion ($3^{\text{rd}}$).\\ 
(B) Top: The spectrum of the solutions in sequence space is similar to that of random sequences (black line), except for about 8-9 high eigenvalues that are outside the continuous spectrum. \\ 
Bottom: the first 8 eigenvectors exhibit patterns
of  alternating +\textbackslash- stripes \---   which we term correlation `ripples' \---  around the fluid channel region. Seeing these ripples through the random evolutionary noise required at least $10^5$ independent solutions \cite{Tesileanu2015}.
}
\end{figure}

\begin{figure}
  \includegraphics[width=\columnwidth]{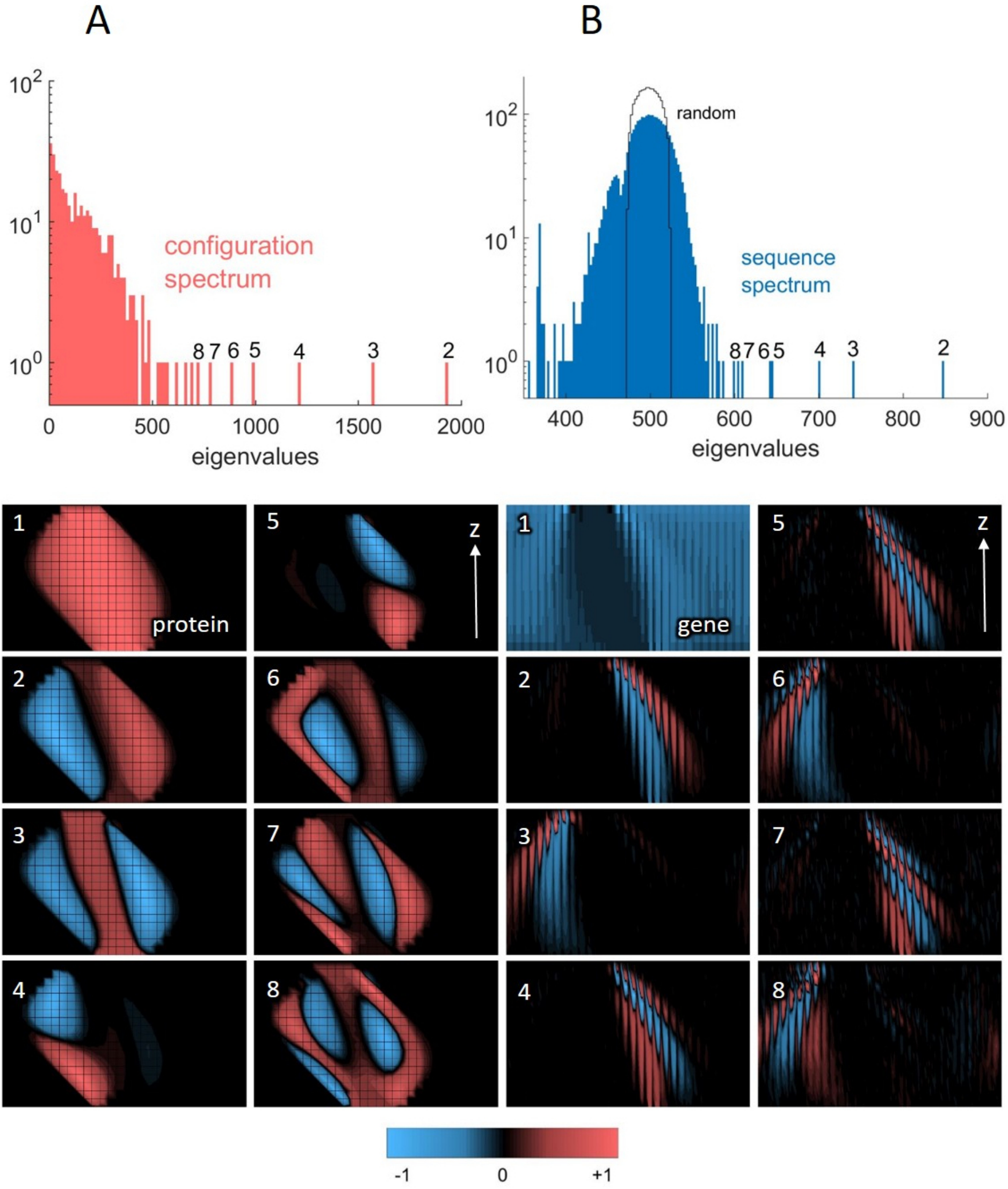}
  \caption{\label{fig:tspec}
  \textbf{Spectra and eigenfunctions for the tilted example (T)}:\\ 
Note the similarity with \fref{fig:fig3}, and also how the tilt is manifested not only in the protein modes , but also in the gene modes. This demonstrates  that the gene and the protein \emph{share}
    common features.\\ 
(A) The configuration spectrum and eigenfunctions.\\ (B) The sequence spectrum and eigenfunctions.
}
  \end{figure}

\subsection{Spectral analysis  reveals  correspondence of genotype  and phenotype spaces} 
\label{sub:spectral}
Spectral analysis of the solution set in both sequence and configuration spaces provides further information on the sequence-to-function map (\fref{fig:fig3}). The sequence spectrum is obtained by singular value decomposition (SVD) of a $10^6 \times 2550$
matrix, whose rows are the binary genes of the solution set. 
The first few eigenvectors (EVs) with the larger eigenvalues capture most of the genetic variation among the solutions, and are therefore the\textit{ collective degrees-of-freedom of protein evolution} (\fref{fig:fig3}B). The $1^\text{st}$ EV is the average sequence, and the next EVs highlight positions in the gene that tend to mutate together to create the fluid channel. 

The spectrum of the
configuration space is calculated in a similar fashion by the SVD of a $10^6 \times 540$ matrix, whose rows are the configurations of the solutions set (\fref{fig:fig3}A). 
In the configuration spectrum, there are 8-10 EVs which stand out from the continuous spectrum, corresponding to the dimension $8$ shown in \fref{fig:fig2}. Although
the dimension of the sequence space is high ($\sim150$), there are again only 8-9 eigenvalues outside the continuous random spectrum.

These isolated EVs distill beautifully the non-random components within the mostly-random functional sequences. The EVs of both sequence and configuration are localized around the interface between the shearable and rigid domains. The similarity in number and in spatial
localization of the EVs reveals the tight correspondence
between the configuration and sequence spaces. 

This duality is the outcome of the sequence-to-function map defined by our simple model: The geometric constraints of forming a shearable band within a rigid shell, required for inducing long-range modes, are mirrored in
long-range correlations among the codons (bits) in sequence space. The corresponding sequence EVs may be viewed as weak 'ripples' of
information over a sea of random sequences, as only about 8 out of 2550 modes are non-random (0.3\%). These information ripples also reflect the self-reference of proteins and DNA via the feedback loops of the cell circuitry \cite{Tlusty2016}.

It is instructive to note similarities and differences between the
spectra.
While the spectra of the
configuration space and of the sequence space have a similar form \textemdash~  with a continuous, more or less random,  part
and a few isolated eigenvalues above it \textemdash~  the \emph{location} of the random part is different: In the configuration case it is close to zero while in the sequence case it is concentrated at large values around $500$. 

The geometric interpretation is that the cloud of solution points
looks like an 8-9 dimensional flat disk in the configuration case,
while in the sequence space, it looks like a high-dimensional
almost-spherical ellipsoid. The  few directions slightly more
pronounced of this ellipsoid correspond to the non-random components
of the sequence. The slight eccentricity of the ellipsoid corresponds
to the weak  non-random signal above the random background. This also illustrates  that the 
dimension of the sequence  space is practically infinite, while in the
configuration space it is comparable to the number of isolated
eigenvalues. 

We verified that the dimensional reduction and the spectral
correspondence depend very little on the details of the models. For
example, we examined a model with $16$ AA species instead of $32$.\footnote{The natural genetic code with its 20 AAs is therefore an intermediate case.} We found that the dimension of the phenotype space was $\sim9.1$, while a lower bound on the genotype dimension was $\sim 150$, very similar to the dimensions of the 32 AA model (compare to \fref{fig:fig2}). The spectra and the eigenmodes of both configuration and sequence spaces were also similar (not shown).

\begin{figure}
\includegraphics[width=\columnwidth]{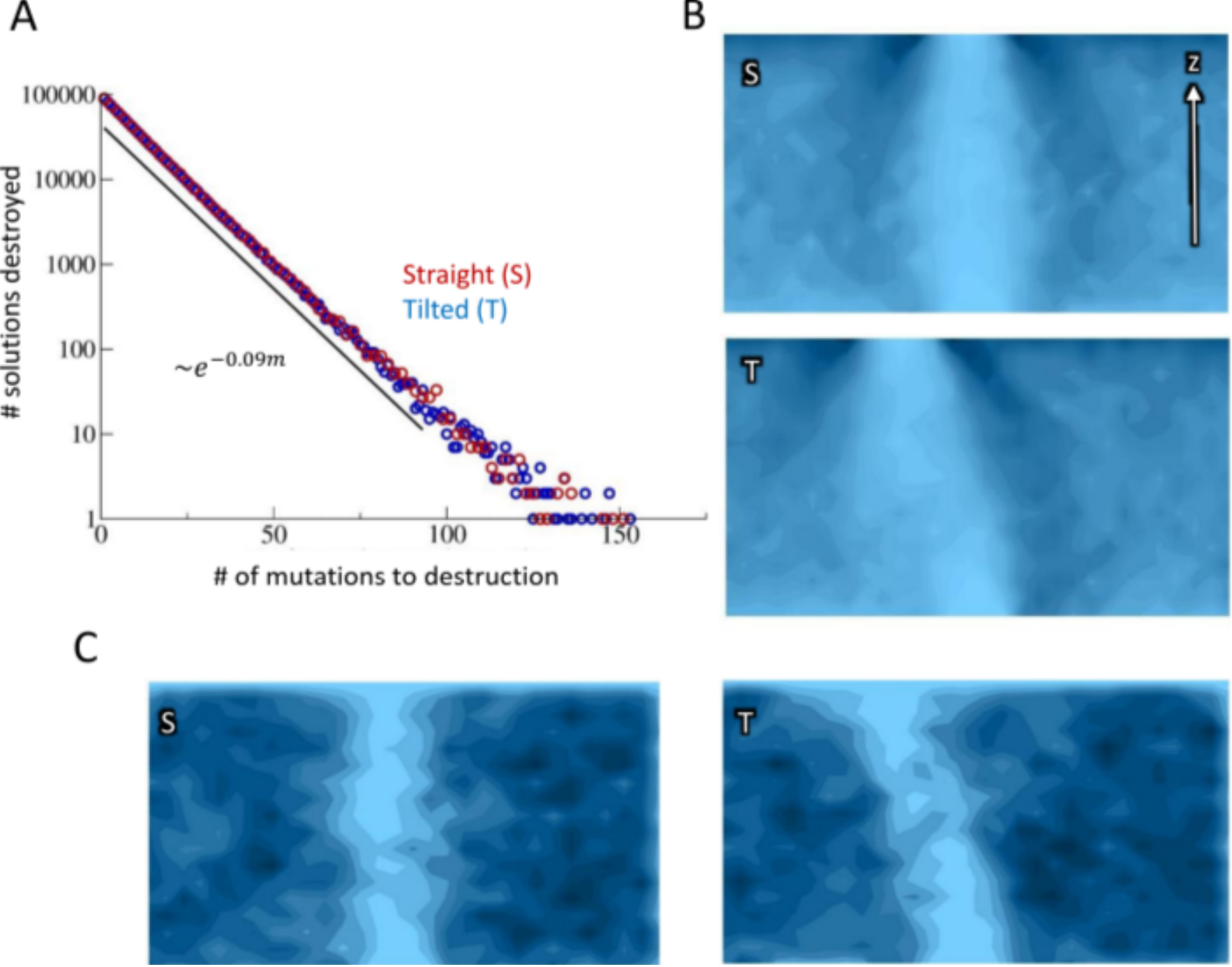}
\caption{\label{fig:fig4}
 \textbf{Stability of the mechanical phenotype to mutations:}\\
  Mutations at sensitive positions of the sequence move the output away from the prescribed solution.\\
(A) Fraction of runs (among $10^6$) destroyed by the  $m$-th mutation. A single mutation destroyed about 9\% of solutions. The proportion  decays exponentially like $\exp({-0.09m})$.\\
(B)  The density map of such mutations for problems S and T (\fref{fig:fig1b}) shows accumulation around the fluid channel and at the top layer (dark
regions).\\
(C) The double mutations are evenly distributed in the rigid regions.}
\end{figure}

\subsection{Stability of the mechanical phenotype under mutations}
\label{sub:stability}
First, we determine how many mutations lead to a destruction
of the solution (\fref{fig:fig4}A).
About $10\%$ of all solutions are destroyed by just one random mutation. The exponentially decaying probability of surviving $m$ mutations signals that these mutations act quite independently.  \fref{fig:fig4}B which shows the location of these destructive mutations around the shearable channel
\footnote{The natural genetic code is redundant, \ie~ several codons encode
the same AA and are therefore synonymous. Such redundancy reduces the fraction of destructive mutations, since mutations that exchange synonymous codons
do not change the encoded AA and are theretofore bound to be neutral. A case
of redundant code is examined in \sref{sub:aa}.}. 

We have also studied the loci where two  \emph{interacting} mutations will destroy a solution (\ie, none of the two is by itself destructive). In most cases, the two mutations are close to
each other, acting on the same site. The channel is less vulnerable to such mutations, but the twin mutations are evenly distributed over the whole rigid network (\fref{fig:fig4}C).

\begin{figure}
\includegraphics[width=0.8\columnwidth]{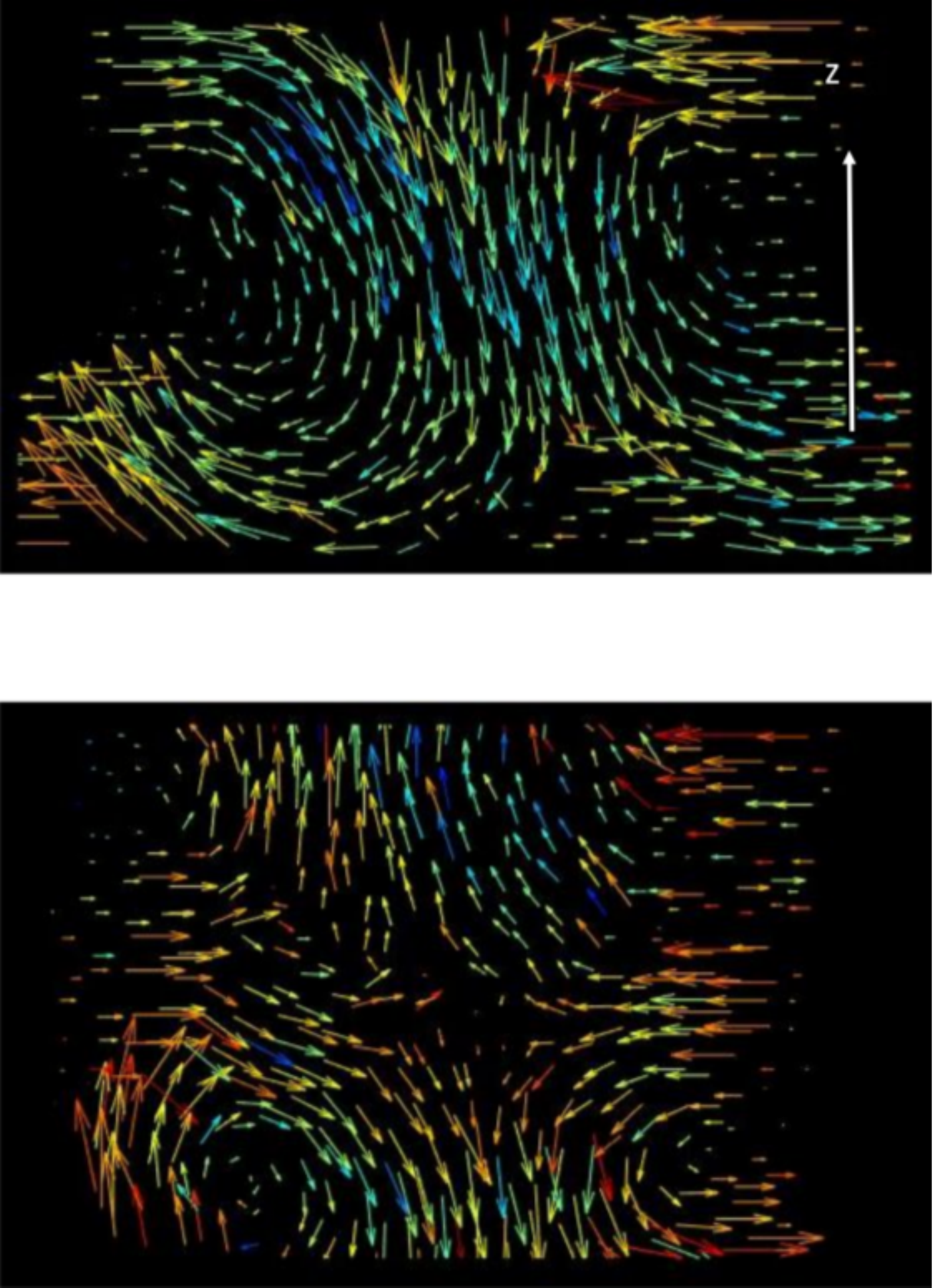}
\caption{\textbf{ Mechanical shear modes:}\\ Displacement and strain fields for the tilted solution T for two low eigenvalues. The vectors show the direction of the displacement and the color code denotes the
strain (\ie, the local change in the vector field as a function of position, maximal stress is red).}\label{fig:fig5}
\end{figure}

\begin{figure}
\includegraphics[width=\columnwidth]{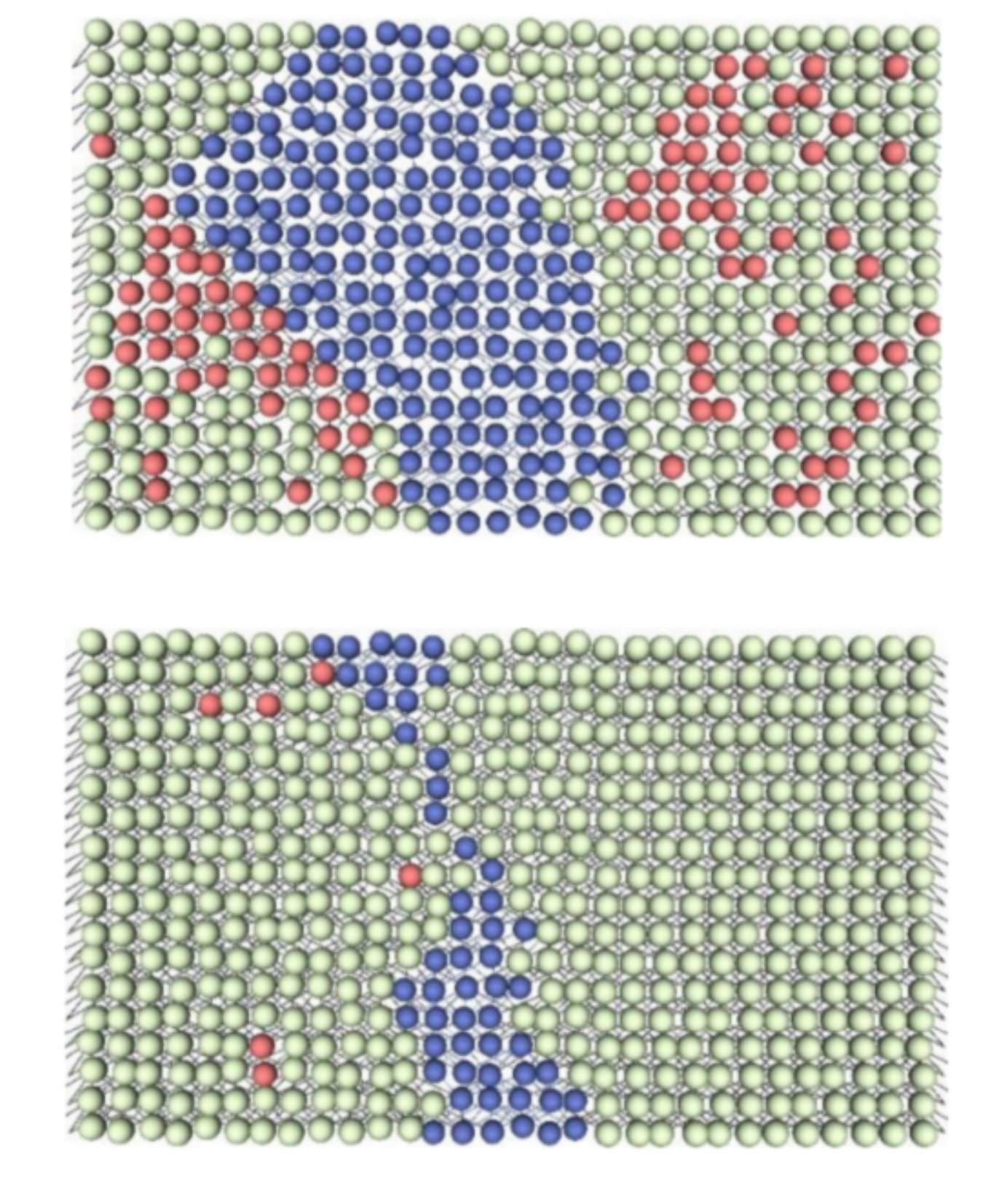}
\caption{
\textbf{Adaptation of thermal stability:}\\ Extreme configurations, with low (50\%, left) and high (95\%,
right) bond density, solve problem T. 
}\label{fig:fig5b}
\end{figure}

\subsection{Fluid channel supports low-energy  shear modes}
\label{sub:fluid}
The evolved rigidity pattern supports low-energy modes with strain localized in the floppy, fluid channel. We tested whether the evolved AA\ network indeed  induces such  modes (\fref{fig:fig5}), by calculating the mechanical spectrum of a spring network in which bonds are substituted by  harmonic springs. The shear motion of the network is characterized by the  modes of $\HH$,  its elastic
tensor. $\HH$ is the $2N$x$2N$ curvature matrix in the harmonic expansion of the elastic energy $E  \simeq\ \half  \mathbf{\delta r}^T \HH~  \mathbf{\delta r}$, where $\mathbf{\delta r}$ is the $2N$-vector of the 2D displacements of the $N$ AAs. $\HH$ has the structure of   the network Laplacian multiplied by the $2$x$2$ tensors of directional derivatives  (see \sref{subs}, which is derived from \cite[pp. 618--9]{Chung1992}).

We traced the mechanical spectrum of the protein during the evolution of the fluidized channel (a shear band). We found that the formation of a continuous channel of less connected amino acids indeed facilitates the emergence of  low-energy modes of shear or hinge deformations  (\fref{fig:fig5}). The energy of such low modes nearly vanishes as the channel is close to completion. Similar deformations, where the strain is localized in a rather narrow channel, occur in real proteins, as shown in recent analysis of structural data  
\cite{Mitchell2016}.

\subsection{Proteins can adapt simultaneously to multiple tasks }
\label{sub:multi}

Our models were designed to trace the evolution  of a mechanical function and show how
it constrains the genotype-to-phenotype map, as shown above.
Real proteins also evolve towards other essential functions, such as
binding affinity and biochemical catalysis at specific binding
sites. Here, we examine another important molecular trait, stability. 

Many studies examine the energetic stability of the protein, as measured by its overall free energy ($\Delta G$) 
\cite{Liberles2012,Zeldovich2008,Koehl2002}. In the present model,
this free energy is given by the number of bonds, which represent chemical and physical interactions among the amino acids. The higher the number of bonds the more stable and less flexible is the protein. By tuning stability, organisms adapt to their environment. Thermophiles that live in hotter places, such as hydrothermal vents, evolve more stable proteins to withstand the heat. Cryophiles that reside in colder niches have more flexible proteins  \cite{Jaenicke1998}. 

We simulated the evolution of the two phenotypes, our specific
dynamical mode together with an energetic state (\ie, a given
bond density).  We find that   the large solution set of the
mechanical problem allows the protein to select  a subset with a specific energetic state. Thus, the evolutionary dynamics could  find solutions to the same
mechanical function when we imposed  extreme values of bond density (\fref{fig:fig5b}). This demonstrates the capacity of the protein to search  in parallel for the solutions of several biological tasks. Evolving a specific binding site is expected to be an easier task, since such sites are confined to a small fraction of the protein. 

\subsection{Amino acid interactions}
\label{sub:aa}

In the model described so far, the bonds were determined by the AA
species alone, while in real proteins, it is the interaction between \emph{pairs} of AA which determines the formation of bonds \footnote{At least two AAs. There may be also higher order terms of three-body interactions etc.}.
This raises the question as to how much our results are sensitive to
the fine details of the interaction model. As
we show, a  more realistic interaction model does not change the main results, which demonstrates  the robustness of our approach.

To model two-body AA interactions we consider a set of three AA species, which
we call $A_0$, $A_1$ and $A_2$. Whether a bond is formed or not is determined
by a symmetric binary relation $b(A_i,A_j) $, which we write as a $3\times3$
interaction matrix,   

\begin{table}[h!]\centering
\begin{tabular}{l|ccc}
&$A_0$&$A_1$&$A_2$\\
\hline
$A_0$& $1$&$1$&$1$\\
$A_1$& $1$&$1$&$0$\\
$A_2$& $1$&$0$&$0$\\
\end{tabular}
  \caption{
  \label{table1}
  The interaction $b(A_i,A_j)$ among the three AAs. The formation of
a  bond by the pair $A_i$-$A_j$ is the denoted by a `$1$', while `$0$' denoted
  the absence of a bond.}
\end{table}

This variant of the  model is reminiscent of the HP model with its two species of AAs \cite{Dill1985}.
The interaction range is kept identical to that our standard model,  
namely an AA can form a bond the 5 nearest neighbors in the adjacent rows. 

The  `gene' in this variant of the model is a sequence of $18 \times 30 = 540$  two-letter binary codons, $g_i$, each representing an AA, such that the overall length
of the gene is $1080$ bits. 
The genetic code is a map $C$ from codons to AAs, $C:g_i \rightarrow A_i$. Since there are four codons and only three AAs, there is a $25\%$ \emph{redundancy} in the `genetic code'. This is  reminiscent of the (higher) redundancy of the natural genetic code in which $20$ AAs are encoded by $61$ codons \cite{Tlusty2007,Eckmann2008,Tlusty2010}
(out of the $4^3=64$ codons $3$ are `stop'\ codons). In our 4-codon genetic code, the redundant AA is chosen to be $A_0$, $C(00)=C(01)=A_0$, and the two other AA are encoded as $C(10)=A_1$, $C(11) = A_2$.
For a given gene, the bond pattern is determined by looking at all AA pairs within the interaction range and calculating their coupling according to the interaction matrix (Table \ref{table1}), $b(C(g_i),C(g_j))=b(A_i,A_j)$.
Once the bond network is determined from the gene, the rigidity pattern, rigid, fluid or `trapped', is calculated as in the standard model 
(\sref{sub:mechanical model}).

In the simulations, at each step we flip one letter in a randomly selected
codon. A quarter of the mutations are synonymous, since they exchange
`$00$' and `$01$'. The other three quarters add or cut bonds, and we
check, as before, whether the connectivity change moves the rigidity pattern
closer to a pattern that allows for a low-energy floppy mode. 
A small number of
beneficial mutations eventually resolve the mechanical transduction
problem, typically after $10^3-10^4$ mutations.  
     

\begin{figure}
\includegraphics[width=\columnwidth]{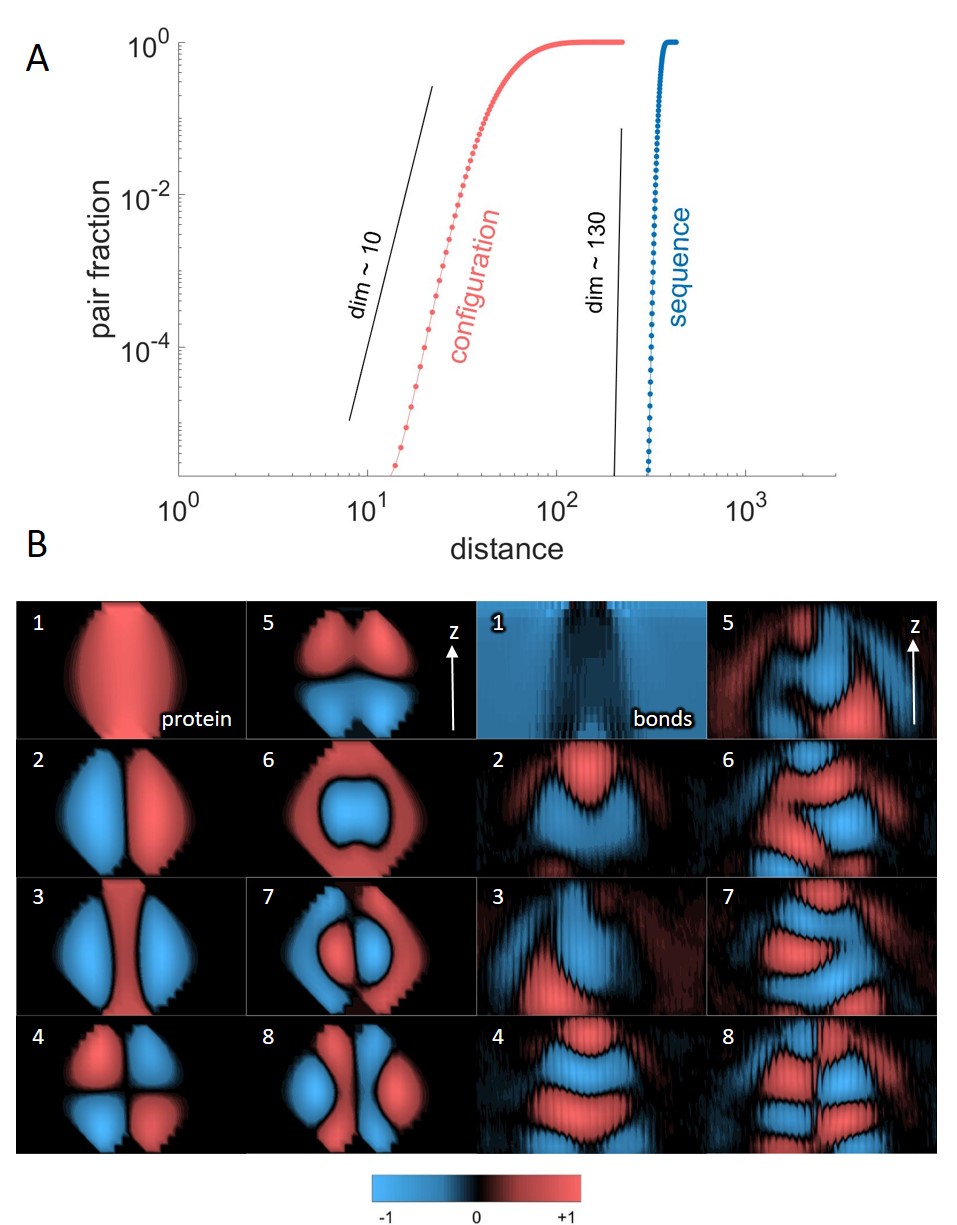}
\caption{\label{fig:aamodel}
(A) {\bf The AA interaction model.} (A) Dimension of the genotype and phenotype spaces are similar to the standard model (\fref{fig:fig2}). (B) Left: The first few eigenfunctions for the configuration. Right: the same for the bond patterns.
 }
\end{figure}

In \fref{fig:aamodel} we present some data (obtained from $4\cdot10^5$
solutions) to illustrate the robustness of the results relative to
model changes.
We find that, despite having changed the connectivity model, our main conclusions regarding the geometry of the phenotype-to-genotype map remain intact: A huge reduction from a high-dimensional genotype space (${\rm dim}> 100$) to a low-dimensional phenotype space (${\rm dim}\sim 10$), similar to the dimensions in \fref{fig:fig2}. It is noteworthy that the configuration eigenvectors are very similar to those of simpler model (as in \fref{fig:fig3}),  although they are determined by very different bonding interactions. This is evident  in  the (non-random) bond eigenvectors which are similar in number to those of the pervious model but differ in pattern owing to the different bonding rules of Table \ref{table1}. The robustness of the results manifests the universality of the dimensional reduction which originates from the continuity of the mechanical transduction.

\section{Conclusions}
\label{sec:conclusions}

Our  models of the genotype-to-phenotype map put forward a new
physical picture  of protein evolution. 
Our thesis is that
rather than structure itself, it is the \textit{dynamics} that governs
protein fitness. 
Our method  considers proteins as evolving  amorphous matter with a mechanical
function, a specific low-energy conformational change. The
rigidity/shearability pattern of the protein, and hence its dynamical
modes, are  determined by the connectivity of the amino acid
interaction network. The model explains how the spatially-extended
modes appear as the gene mutates and changes the amino
acid network. 
These modes are shear and hinge motions where the strain is localized in the shearable channel and where the surrounding domains translate or
rotate as rigid bodies (\fref{fig:fig5}). 

A  main insight from our model is that
requiring the protein to have `floppy' modes puts strong constraints
on the space of mechanical phenotypes. 
As a consequence there is a huge dimensional reduction when mapping
genotypes to phenotypes. We find that the collective mechanical
interactions among the amino acids are mirrored in corresponding modes
of sequence correlation in the genes.  These main results do not
depend on details of the model and have been reproduced in versions
with (i) a different number of AA species (16 instead of 32), (ii)
bonds that depend on pairwise interactions, and (iii) harmonic  spring
network \cite{Dutta2017}. All these suggest that the results are
generic and apply to a wide range of realizations.

Our models are distilled to their simplest physical-mathematical
schemes, but have concrete,
experimentally testable predictions. 
In the functional protein, the least random, 
strongly correlated sites are concentrated in a rigid shell that envelops the shearable channel \cite{Mitchell2016}. Our
model therefore predicts that these sites are also the most vulnerable to mutations (\fref{fig:fig4}B), which distort the low-frequency modes and thus hamper the biological function.
These effects can be examined by combining mutation surveys, biochemical assays of the function, and physical measurements of the low-frequency spectrum, especially in allosteric proteins. 

To that end, one may take an enzyme with a known shear band (via
analysis similar to \cite{Mitchell2016}) and mutate amino acids within
and around the band. We expect the mutation of these  amino acids to
have a significant impact on the dynamics and biochemical function of
the protein, as compared to other mutations in the rigid
subdomains. By sequence alignment methods
\cite{Casari1995,Gogos2000,Tesileanu2015,Rivoire2016}, it is possible
to test whether these sensitive positions in the protein exhibit
strong correlations in the gene, as predicted by the model. One may
also  search for the dimensional reduction predicted by the model in
high resolution maps of molecular fitness landscapes
\cite{Jacquier2013,Roscoe2013,Firnberg2014,Kondrashov2015}. 

Past studies have shown that the motion of proteins \cite{Levitt1985,Tirion1993,Bahar2005,Zheng2006} and their hydrophobicity patterns \cite{England2011} may often be approximated by a few normal modes, while others have demonstrated that the variation in aligned sequences may be characterized by a few correlation modes
 \cite{Casari1995,Gogos2000,Tesileanu2015,Rivoire2016}. The present
 study links the genotype and phenotype spaces, and explains the
 dimensional reduction as the outcome of a non-linear mapping between 
genes and  patterns of mechanical forces: We characterize the emergent functional mode to be a soft, `floppy' mode, localized around a fluidized channel (a shear band), a region of lower connectivity which
is therefore easier to deform.  The contiguity of this rigidity pattern implies that it can be described by a few collective degrees of freedom,  implying a vast dimensional reduction of configuration space.   

The concrete genotype-to-phenotype map in our simple models demonstrates that most of the gene records random evolution, while only a small non-random fraction is constrained by the biophysical function. 
This drastic dimensional reduction is the origin  of the flexibility and evolvability  in the functional solution set.

\appendix

\section{The protein evolution model}

\subsection{The  cylindrical amino acid network}

We model the protein as an aggregate of amino acids (AAs) with short
range interactions. In our coarse grained model, beads represent the
AAs and bonds their  interactions with neighboring AAs (\fref{fig:fig1}). We
consider a simplified cylindrical geometry, where the AAs are layered
on the surface of a cylinder at randomized positions, to represent the
non-crystalline packing of this amorphous matter. Throughout this study,
we examine a geometry with height $h(=18)$, \ie, the
number of layers in the $z$ direction, and width $w(=30)$, \ie,
the circumference of the cylinder. When the cylinder is shown as a
flat 2D surface 
(such as in \fref{fig:fig1b}), there are still periodic boundary conditions in the horizontal $w$ direction.
 The row and column coordinates of an AA are $(r,c)$, with $r$ for the
 row $(1,\dots,h)$ and $c$ for the
column $(1,\dots,w)$.
The cylindrical periodicity is accounted for by taking the horizontal coordinate $c$ modulo $w=30$, $c \rightarrow \mod_w(c-1) + 1$.

Each AA in row $r$  can connect to any of its five nearest neighbors
in the next row below, $r-1$. This defines $2^5 = 32$ effective
species of amino acids that differ by their `chemistry', \ie, by the
pattern of their bonds. Therefore, in the gene, each AA at $(r,c)$ is
encoded as a $5$-letter binary \textit{codon}, $\ell_{rck}$, where the \textit{k}-th letter denotes the
existence ($= 1$)\ or absence ($ = 0$) of the \textit{k}-th  bond. The
gene is the sequence of  $N_{AA} = w\cdot h=540$ codons which
represent the AAs of the protein. This means that each codon just
specifies which of the 5 bonds are present or absent.
Therefore, the codons are a genetic  \emphx{sequence} of
$2700 = w\cdot h\cdot 5$ digits 0 or 1. Each of these numbers
determines 
whether or not a \emphx{bond} connects two positions of the grid. Since the bonds from the bottom row do not affect the configuration of the protein and the resulting dynamical modes, the relevant length of the gene is somewhat smaller,
$N_S=2550 = w\cdot (h - 1)\cdot 5$. 

\subsection{Evolution searches for a mechanical function }

We now \emph{define} the target of evolution as finding a functional
protein, in the following specific sense:
To become functional, the protein has to evolve a \emph{configuration} of AAs and
\emph{bonds} that can
\emph{transduce a mechanical signal} from a prescribed input at the bottom of
the cylinder to a prescribed output at its top. This signal is a
large-scale, low-energy deformation where one domain moves rigidly
with respect to another in a shear or hinge motion, which is
facilitated by the presence of a fluidized, `floppy' channel
separating the rigid domains \cite{Alexander1983,Phillips1985,Alexander1998}. 

\subsection{Rigidity propagation algorithm }\label{sec:rigidity}

\begin{figure}[h]
  \includegraphics[width=\columnwidth]{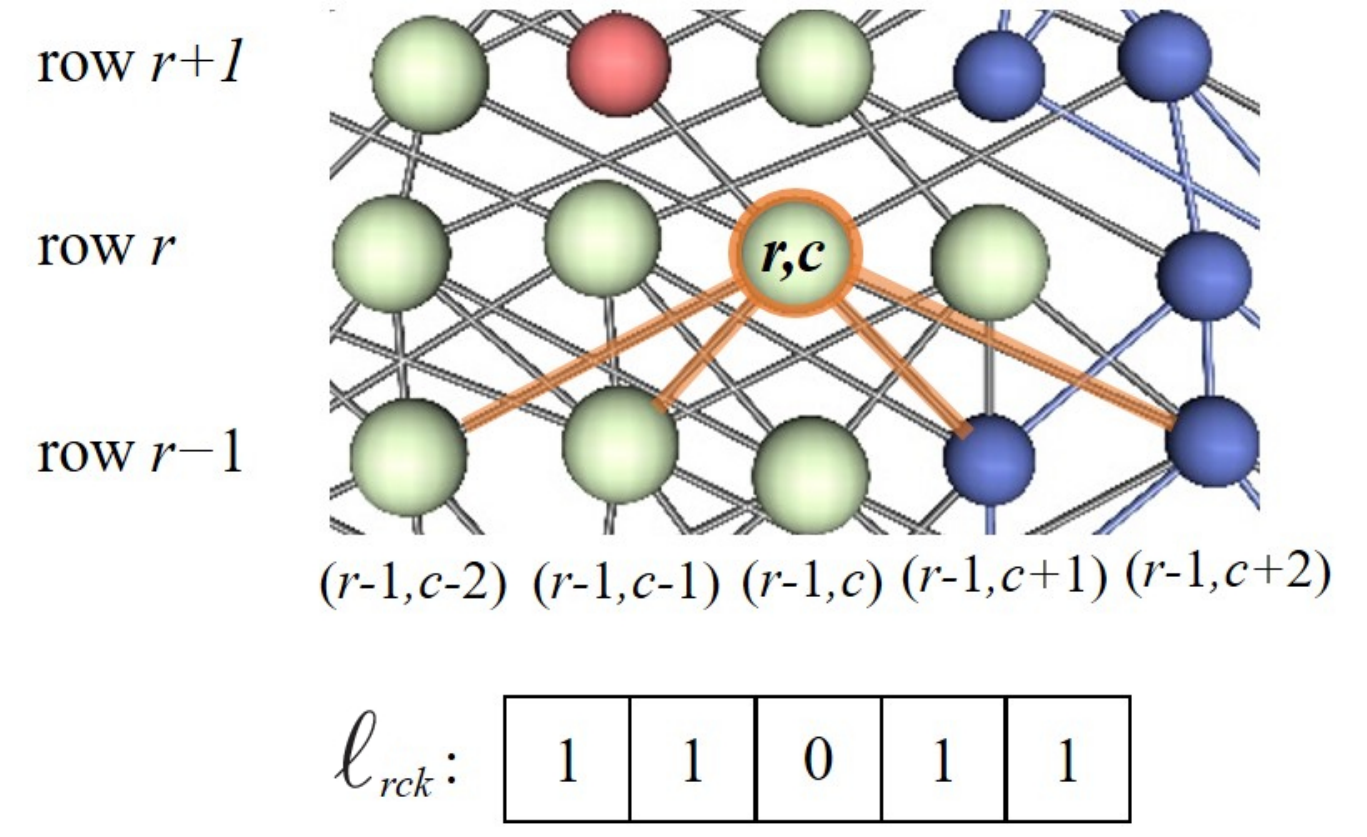}
  \caption{\label{fig:st}Illustration of the percolation rules for shearability and fluid/solid states. Note that site $(r,c)$ was turned solid because it is attached to 2 solid sites below it. Also note that the
  red site above it is fluid, because it is attached to less than 2 solid sites below it. But it is not shearable because it does not connect to a shearable site below it. On the other hand, the top right site is shearable and fluid, since it is attached to only one   solid site (namely $(r,c)$) and no others on the invisible part of the structure (as seen by its blue connections), and it is also   connected to the blue site at $(r,c+2)$.}
\end{figure}

The large-scale deformations are governed by the rigidity pattern of the configuration, which is determined by the connectivity of the AA network via a simple majority
rule (\fref{fig:fig1}).
The details of this majority rule are as follows 
(\fref{fig:st}): Each AA position will have two binary properties, which define its state:
\begin{myitem}
\item The \emphx{rigidity} $\sigma$: This property can be \emphx{rigid} ($\sigma=1$) or \emphx{fluid} ($\sigma=0$).
\item The \emphx{shearability} $s$: This property can be \emphx{shearable} ($s=1$) or \emphx{non-shearable} ($s=0$). As shown below, a non-shearable AA can be either rigid or fluid within a rigid domain of the protein. Non-shearable domains tend to move as a rigid body (\ie, via translation or rotation), whereas shearable regions are easy to deform. 
\end{myitem}

Only 3 of the 4 possible combinations are allowed  :
\begin{myenum}
  \item Non-shearable and solid AA (yellow): ($\sigma = 1; s = 0$).
   \item Non-shearable and fluid AA (red): ($\sigma = 0; s = 0$).
     \item Shearable and fluid AA (blue): ($\sigma = 0; s = 1$).
     \item Shearable solid is forbidden.
\end{myenum}

Given a fixed sequence, and an \textit{input} state in the bottom row of the
cylinder, $\{\sigma_{1,c}$, $s_{1,c} \}$ the state of the cylinder is completely determined   as follows:
The three states  percolate through the
network, from row $r$ to row $r+1$ (see \fref{fig:st}). This propagation is directed by the
presence of bonds, with a maximum of $5$ bonds ending in each AA (of rows $r = 2$ to $h$; the state of the first row is given as input).
These bonds can be \emph{present}(=1) or \emph{absent}(=0).
according to the codon $\ell_{rck}$, $k=-2,\dots,2$ when they point to the AA
with coordinate $(r,c)$ coming from the AA $(r -1,c+k)$. 

In a first sweep through the rows, we deal with the \emph{rigidity} property $\sigma$.
In row $r = 1$ each of the $w$ AAs is in a rigidity state rigid
($\sigma=1$) or fluid ($\sigma=0$). In all other rows, $r = 2$ to $h$, the 5 bonds determine the value of
the rigidity of $(r,c)$  through a majority rule:
\begin{equation}\label{eq:1}
\sigma_{r,c}= \theta \left( \sum_{k=-2}^2 \ell_{rck} \sigma_{r-1,c+k} - \sigma_0 \right),
\end{equation}
 where $\theta$ is the step function ($\theta(x \ge 0) =1$, $\theta(x<0)=0$)). The parameter  $\sigma_0 = 2$ is the minimum number of rigid AAs from the $r-1$ row that are required to rigidly support AA: In 2D each AA\ has two coordinates which are constrained if it is connected to two or more static AAs.
In this way, the rigidity property of being pinned in place propagates through the lattice, as a function of the initial row and the choice of the
bonds which are present as encoded in the gene. 

We next address the \emph{shearability} property. It is
determined by the rigidity of  AAs as follows: We assume
that all fluid AAs in row $r=1$ are also shearable (blue: ($\sigma = 0; s = 1$)). A fluid  node $(r,c)$ in row $r$
will become shearable exactly if at least one of its neighbors $(r-1,c)$ or
$(r-1,c\pm1)$ is shearable:
\begin{equation}\label{eq:2}
s_{r,c}= (1-\sigma_{r,c})\cdot\theta \left( \sum_{k=-1}^1 {s_{r-1,c+k}} - s_0 \right),
\end{equation}
where $s_0 = 1$. The first term on the lhs ensures that a solid AA can never become shearable.
This completes the definition of the map from the sequence to the state.

\subsection{Fitness and mutations}

As we explained before, the aim is to find a functional protein which
can transfer forces. To find such a protein, we start from a random
sequences (of 2550 codons), and from an initial state (\emph{input}) in the bottom
row of the cylinder. This initial state is just made from rigid and
fluid beads, as shown \eg, in \fref{fig:fig1b}. For most simulations,
we just took 5 consecutive fluid beads among the remaining solid beads.

We next define the \emphx{target}. It is a chain of $w$ values,
fluid and shearable ($\sigma = 0; s = 1$) or solid ($\sigma = 1; s = 0$), in the top row, which the protein should yield as an \textit{output}: $\{\sigma^*_{c}$, $s^*_{c} \}_{c=1,\dots,w}$.
Given (i) a gene sequence, which determines the connectivity
$\ell_{rck}$  and (ii)\ the \textit{input} state, $\{\sigma_{1,c}$,
$s_{1,c} \}_{c=1,\dots,w}$, the algorithm described above uniquely defines the
output state in the top row, $\{\sigma_{h,c}$, $s_{h,c} \}_{c=1,\dots,w}$. At each
step of evolution, the output state is compared to the fixed, given target, by
measuring the Hamming distance, the number of positions where the output differs from the target:
\begin{equation}\label{eq:3}
F = \sum_{c=1}^w{\left[ 1-\left(\lvert s_{h,c} - s_{c}^* \rvert -1\right)\cdot
\left(\lvert \sigma_{h,c} - \sigma_{c}^* \rvert -1 \right) \right]}.
\end{equation}
In the biological convention $-F$\ is the \textit{fitness} that should
increase towards a maximum value of $-F=0$, when the input-output
problem  is \emphx{solved}.

Solutions are found by \emph{mutations}. At each iteration, a randomly
drawn digit in the gene is flipped, that is, the values of 0 and 1 are
exchanged. This corresponds  to erasing or creating  a randomly
chosen  link of a randomly chosen AA. After
each flip, a sweep is performed, and the new output at the top row is again compared
to the target. A mutation is kept \emph{only if the Hamming distance
  is not increased as compared to the value before the mutation} (equivalently the fitness is not allowed to decrease). 
This procedure is repeated until a solution $(F=0)$ is found. This will
happen with probability 1, perhaps after very many flips, if the
problem has a solution at all. This is really the Metropolis algorithm \cite{Metropolis1953} algorithm
(at 0 temperature).

{\bf Remark}: {\sl It is an important feature of our model that the quality of a
  network is \emph{only} measured at the target line. This corresponds
to the biological fact that the protein can only interact with the
outside world through is surface (in our case, the ends of the
cylinder). One of the surprising outcomes of our study is that this
requirement has a strong influence on what happens in the interior of
the protein. Also, the propagation of fluidity should not be confused
with learning in neural networks, but is rather of the percolation type.}

\subsection{Simulation of evolutionary dynamics}

All simulations are done on the $30\times 18=540$ playground, as described
above. We have done simulations for many variants of the model, and
many targets, but we present only two specific problems, for which the
most extensive study was done: In the first, the
fluid regions of the input and the target are opposite and of length 6
at the bottom and length 5 at the top. In the second run, top and
bottom are the same, but the top is shifted sideways by 5 units. We
will call these two examples \emphx{straight} and \emphx{tilted},
denoted as S and T. We have also studied examples in which the
position of the target (relative to the input) is left free, but here
we only discuss the results for the 'S' and 'T' case. This serves to
illustrate that the results are largely independent of the details of
the model. We have studied many other variants, and in all cases, the
main results are qualitatively unchanged.

{\bf Remark}: {\sl We view this as an important outcome of our theory,
  namely that it illustrates a close connection between gene and
  protein which goes way beyond the simple model we consider here.
}

For both, S and T, we study 200 independent \emph{branches}\index{branch}, starting from a
random sequence with about $90\%$ of the bonds present at the
start. Given any fixed sequence, we sweep according to the rules
of \eref{eq:1}-(\ref{eq:2}) through the net, and measure the Hamming distance
$F$ (\eref{eq:3}) between the last row and the desired target. 
When this Hamming distance is 0, we consider the problem as solved.
If not, we flip randomly a bond (exchanging 0 with 1) and recalculate
the Hamming distance. We view this flip as a \emphx{mutation} of the
sequence, equivalent to mutating one nucleic base in a gene. 
If the Hamming distance decreases or remains unchanged, we keep the
flip, otherwise we backtrack and flip another randomly chosen bond. 
This is
repeated until a solution is found. (This is really a Metropolis algorithm \cite{Metropolis1953} at zero temperature.) Typically, after $10^3$-$10^5$ mutations this
input-output problem is solved. 
Although the functional
sequences are extremely sparse among the $2^{2550}$ possible sequences, the small bias for getting closer to the
target in configuration space directs the search
rather quickly.

Once a solution is found, we destroy it by further mutations and then
look for a new solution, as before, starting from the destroyed
state. This we call a \emphx{generation}. For each of the 200 branches, we followed 5000 generations, leading to a total of $10^6$
solutions.
The time to recover from a destroyed state is about 1500 flips per
error in that state, which is similar to time it takes to find a
solution starting from a random gene. A  destruction takes around 11.2
mutations on average.

We also did another $10^6$ simulations starting each time from another
random configuration. The statistics in both cases are very similar,
but the destruction-reconstruction simulations obviously show some
correlations between a generation and the next. This effect disappears
after about 4 generations.

\section{Results, analysis  and interpretation}

\subsection{Dimension of solution set}\label{sec:dimension}

Dimension of a space measures the number of directions in which one
can move from a point. In the case of our model, since from any
sequence in sequence space
one can move along $N_S=2550$ axes by flipping just one bit,
we see that the sequence space  has dimension 2550, and the
number of different elements in this space is a hypercube with
$2^{2550}\sim 10^{768}$ elements.

The set of solutions which we find, has however much smaller dimension, as we show in \fref{fig:fig2} for the straight and tilted example. In the 
case of experimental data, as ours, the dimension is most conveniently determined by the box-counting (Grassberger-Procaccia
\cite{Grassberger1983}) algorithm. This is obtained by just counting the number
$N(\rho)$ of pairs at distances $\le \rho$, and then finding the slope
in a log-log plot. This is indicated by the black lines in
\fref{fig:fig2}  we see that, clearly, the dimension in the
space of configurations is about 8-9, while, in the space of
sequences, the dimension is basically `infinite', namely just
limited by the maximal slope one can obtain \cite{Procaccia1988}.

\subsection{Spectrum in phenotype and genotype spaces}\label{sec:spectrum}
We compute spectra for both the sequences and the
configurations, for the $10^6$ solutions. Let us detail this for the
case of sequences: We have $10^6$ binary vectors with $N_S=2550$
components each, and we want to know the `typical' spectrum of such
vectors. This is conveniently found with the Singular Value
Decomposition (SVD), in which one forms a matrix $W$ of size $m\times
n=10^6\times 2550$. This matrix can be written as $U\cdot D\cdot V^*$,
where $U$ is $m\times m$, $V$ is $n\times n$ and $D$ is an 
$m\times n$ matrix which is diagonal in the sense that only the
elements $D_{ii}$ with $i=1,\dots,n$ are nonzero. (We assume here that
we are in the case $m>n$.) The $D_{ii}$ are in general $>0$ and in this
case the singular value decomposition is unique. We call the set of
the $\lambda^{G}_i = D_{ii}$ the spectrum of the sequences, and the vectors in $V$
the eigenvectors of the SVD. It is the first
few of those which are shown in \fref{fig:fig3}.

Note that the SVD\ eigenvalues $\lambda^{G}_i$ are the square roots of the spectrum of the covariance matrix $W^{\rm T}W$ which has the same eigenvectors as $W$. Therefore the high SVD eigenvalues correspond to the \emph{principal components}, the directions with maximal variation in the solution set.     

Mutatis mutandis, we perform the same SVD for the case of the
configurations, using the $s$-values (that is, of the shearability) of
vectors of the configurations. (This is reasonable, because, in
general, there are very few non-shearable and fluid AAs.)

Apart from the numerical findings, which are shown in \fref{fig:fig3} for the
straight (S) example and in \fref{fig:tspec} for the tilted (T) one,
some comments are in order:

\textbf{Configuration space} (The eight figures on the bottom left): The first mode is proportional to the
average configuration. The next modes reflect the basic deviations of
the solution around this average. For example, the second modes is
left-to-right shift, the third mode is expansion-contraction
etc. Since, the shearable/non-shearable interface can
move at most one AA sideways between consecutive rows, the modes are
constrained to diamond-shaped areas in the center of the
protein. This is the joint effect of the `influence zones' of the input and
output rows. 

\textbf{Sequence space} (The eight figures on the bottom right): The first
eigenvector is the average bond occupancy in the $10^6$ solutions. The
higher eigenvalues reflect the structure in the many-body correlations
among the bonds. The typical pattern is that of `diffraction' or
`oscillations' around the fluid channel. This pattern mirrors the
biophysical constraint of constructing a rigid shell around the
shearable region. Higher modes exhibit more stripes, until they become
noisy, after about the tenth eigenvalue. 

The bond-spectrum, top right
in  Figs. \ref{fig:fig3} and \ref{fig:tspec}, has some outliers, which correspond to the
localized modes shown in the eight panels below. Apart from that, the
majority of the eigenvalues seem to obey the Mar\v cenko-Pastur
formula, see \cite{Marcenko1967}. If the matrix is $m\times n$, $m>n$,
then
the support of the spectrum is $\frac{1}{2}(\sqrt{m}\pm\sqrt{n})$. In
our case, since we have a $10^6\times 2550$ matrix, one expects (if they were really random) to find the
spectrum at $\frac{1}{2}(\sqrt{10^6}\pm \sqrt{2550})$, which is close
to the experiment, and confirms that most of the bonds are just
randomly present or absent. We attribute the slight enlargement of the
spectrum to memory effects between generation in the same branch. This
corresponds to the well-known phylogenetic correlations among
descendants in the same tree.

It is tempting to also study the continuous part of
this spectrum, which is not quite of the standard form. While in
principle, this could be done by taking into account the known
correlations, even the techniques of \cite{Guhr1998} seem
difficult to implement. We thank T.~Guhr for helpful discussions on
his issue.

\subsection{Shear modes in the amino acid\ network}\label{subs}
 
Consider now either of the two examples, straight or tilted (S and T).
A solution of such an example is given by a set of bonds, and this set of
bonds defines a graph on the $N_{AA}=h\cdot w=540$ AAs. This graph is
embedded in 2D where $\vec{ x}_{r,c}$ are the positions of the AAs,
which are connected by straight bonds. We now extend the scope of our
study somewhat, by assuming that the bonds are not totally rigid, but
given by harmonic springs (see also \cite{Dutta2017}). This allows us
  to study mechanical properties which would be too stiff if we only
  worked with bonds which are rigid sticks.

In this case, the calculations are straightforward, if somewhat
complex, and they are, \eg, well explained in \cite[pp. 618--619]{Chung1992}.
We thus consider the elastic
tensor, $\HH$,
which is the tensor product of the network Laplacian with the 2 by 2
tensor of directional derivatives.

For the reader who is unfamiliar with \cite{Chung1992}, we describe what this means component-wise. 
The playground $\Omega\subset {\bf Z}^2$ has size $h$ in the
$z$-direction and size $w$ in the $x$ direction, with periodic
boundary condition in the $x$ direction.
All bonds go from some $(r,c)$ to $(r+1,c)$, $(r+1,c\pm1)$,
$(r+1,c\pm2)$, again with periodic boundary conditions in the $c$-direction.
Each such bond defines a direction vector $(d_z,d_x)$ in ${\bf R}^2$ which we
normalize to $d_x^2+d_z^2=1$. Note that this vector depends on both
the origin and the target of the bond.

If we imagine harmonic springs between the nodes connected by bonds
(all with the same spring constant), then we can define the
(symmetric) tensor matrix of
deformation energies in the $x$ and $y$ direction by
\begin{equation*}
  \HH'_{km}= M(k,m)~, \text{with } k,m \in \Omega~,
\end{equation*}
and where each element of $\HH_{km }'$ is---when $k$ and $m$ are connected
by a bond---the 2 by 2 matrix (indexed by $i,j\in\{1,2\}$)
\begin{equa}
  M(k,m )&=\left(d_x(k,m),d_z(k,m)\right)^{\rm T}\otimes \left(d_x(k,m),d_z(k,m)\right)\\&=
  \begin{pmatrix}
    d_x^2&d_xd_z\\
    d_xd_z&d_z^2\\
  \end{pmatrix}~.
\end{equa}
If $k$ and $m$ are not connected, then $M(k,m )$ is the 0
matrix. The elements of $M(k,m)$ are denoted $M(k,m)_{ij}$.

Finally we complete the $2N\times 2N$ matrix $\HH'$ to a `Laplacian'
$\HH$ by adding diagonal elements to it, so that the row (and column)
sums are 0. In components, this means that we require, for each
$k\in \Omega$ 
and each $i,j\in\{1,2\}$, the sums 
\begin{equation*}
  \sum_{\ell} \left( \HH_{km} \right)_{ij}
\end{equation*}
to vanish. 
Other properties of $A$ are described in \cite{Chung1992}.

Since we take periodic boundary conditions in the $x$ direction,
there will always be a (simple) 0 eigenvalue of $\HH$ in this direction. Other 0 eigenvalues correspond to translation in the $z$ direction or rotation in the $x-z$ plane. Another type of 
(double) 0 eigenvalues are associated with any patch of nodes which
is totally disconnected from the rest of the lattice. Since the
density $\rho$ of bonds is about $1/2$ and otherwise quite random, and
there are twice 5 bonds at each interior node we expect (assuming
random distribution of bonds)
there to be about $N\cdot2^{-10}\sim 0.001N$ isolated nodes, \ie, isolated
singletons, and even fewer patches of greater size.

Further zero modes come from 
nodes which can oscillate sideways without first order effects. This
will happen if a node is only connected by one bond.
Since $\rho\sim 1/2$, the probability of finding such a node is about
\begin{equation*}
N\frac{
  \binom{10}{1}}{2^{10}}\sim 0.01 N~.
\end{equation*}
Thus, we show in Figures \ref{fig:fig3} and \ref{fig:tspec} 
 the eigenfunctions only for the first eigenvalues
after the trivial ones. Due to the tensorial nature of the problem,
the eigenvectors have two components, which we show as
2D shear-flow.

\subsection{Genetic correlation matrix}\label{sec:correlation} 
In \fref{fig:fig2c}, we study the correlations among the
$10^6$ solutions in sequence space. Given the matrix $W_{ij}$, of all
sequences, with
$i=1,\dots, N=10^6$, $j=1,\dots,2550$ (of binary digits), we compute
the means
$\langle W_{\cdot\, j}\rangle=\sum_{i=1}^N W_{ij}/N$  and the standard deviations
$\text{std}_j=\left(\sum_i |W_{ij}-\langle
W_{\cdot\, j}\rangle|^2\right)^{1/2}$. Then, in the usual way, we form
$M_{ij}=W_{ij}-\langle W_{\cdot\, j}\rangle$ and 
$$
C_{j,j'}=\frac{(M^* M)_{j,j'} }{\text{std}_j \text{std}_{j'}}~.
$$
\fref{fig:fig2c} then shows $\log(|C_{j,j'}|)$, with the autocorrelation
$C_{jj}$ omitted.

Note that both, the means and the variances depend very weakly on $j$. \fref{fig:fig2c} reveals and reinforces several observations also made in
other calculations of this paper. First, looking onto the axis $j=j'$
in the figure one sees a periodicity of the patterns corresponding to
the 17 gaps between the 18 rows of the configuration space.
This reflects the necessity to maintain a \emph{connected} liquid channel.
Also, as seen in \fref{fig:fig2c}, the
correlations grow somewhat towards the ends, especially toward the
upper ($j=2550$) end. This is because of the mechanical constraint
which forces the channel to become more precise towards the ends, in
analogy with \fref{fig:fig4}B.

The periodic patterns all over the square reflect not only the natural
periodicity of 150 ($=5\cdot w$) elements in the sequence, but also show that the
boundaries of the channel form a special shell (with \emph{two} peaks
per row).

\subsection{Survival under mutations}
Here, we ask how robust the solutions are as further mutations take
place. First, we determine how many mutations lead to a destruction
of the solution. The statistics of this is shown in
\fref{fig:fig4}.
We note that about $10\%$ of all solutions are destroyed by just one
mutation, while there is an exponential decay of survival of $m$
mutations.
This signals that the mutations act independently. 

One can also ask \emph{where} the critical mutations take
place.  This
is illustrated in \fref{fig:fig4}B, and was discussed in the main text.
We have also studied the places where exactly \emph{two} mutations will kill a
solution (and none of the 2 is a single site `killer')
(\fref{fig:fig4}C) and in
these cases, one finds that the two mutations are generally close to
each other, acting on the same site. Again, the channel is less
vulnerable to mutations but now the mutations are evenly distributed
over the rest of the network.

\subsection{Expansion of the protein universe}
Let us explain in further detail how \fref{fig:fig2b} was obtained.
Here, we test our model against the ideas of \cite{Povolotskaya2010}. Our results
will give some insight about the nature of the graph of solutions.
First, we describe the question as it is found in \cite{Povolotskaya2010}. Take any
two solutions and consider their gene sequences $s_1$ and $s_2$. They
will have a Hamming distance $d(s_1,s_2)$, which we normalize by
dividing by 2550 (the number of elements in $s_i, i=1,2$), which we call the protein universe diameter. The
question is how much the solution following one generation after $s_2$
differs from $s_1$. If we call that solution $s_3$, then the observed
quantity is defined as follows: Let $w_i=1$ if $s_{1,i}=1$ and $-1$ if
$s_{1,i}=0$, for $i=1,\dots 2550$.
Then for each $i$ let $x_i=w_i\cdot (s_{3,i}-s_{2,i})$. Note that
$x_i>0$ if the change between $s_{3,i}$ and $s_{3,i}$ is
\emph{towards} $s_{1}$ and $<0$ if it is \emph{away from} $s_1$.
Finally, $N_{\text{away}}=\sum_{i:w_i<0}1$ and
$N_{\text{towards}}=\sum_{i:w_i>0}1$, and we plot in \fref{fig:fig2b}
$N_{\text{towards}}/N_{\text{away}}$ as a function of $D$.

In \fref{fig:fig2b} we show the results for data set S, (the plot for
set T looks similar).
The black curve is nothing but $D/(1-D)$, where $D$ is the normalized
Hamming distance, \ie, the proportion of sites which are different
between $s_1$ and $s_2$. The fit to this curve tells us an important aspect
about the set of possible solutions. Note that the set of all possible
$s$ forms a hypercube of dimension $2550$ with $2^{2550}$ corners. The
set of solutions is a very small subset of this hypercube, where all
corners which are not solutions have been taken away, including the
bonds leading to these corners. This leads to a very complicated sub-graph
of the hypercube. While we do not have a good mathematical description of
how it looks, the good fit shows that the comparisons between $s_1$,
$s_2$, and $s_3$ are \emph{as if  one performed a random walk on the full cube}.
(Note that such a result must be intimately connected to the high
dimension of the problem, since for low dimensional hypercubes it
does not hold.) Almost all solutions are at the edge of the universe, where the typical Hamming distances among the sequences are close to the typical distance between random sequences,

\subsection{Flexibility of solutions: thermal stability}

\begin{figure}
  \includegraphics[width=0.8\columnwidth]{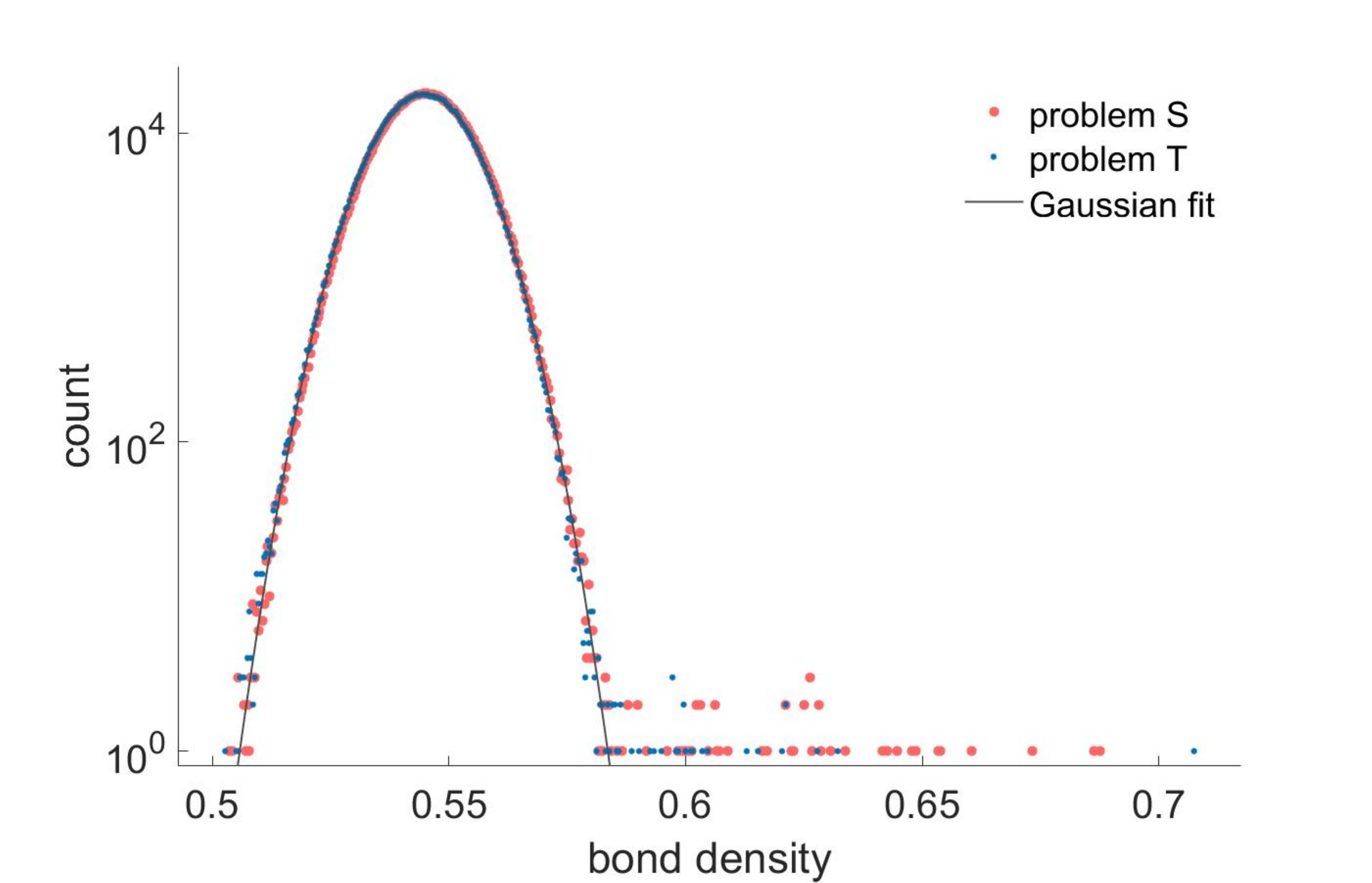}
  \caption{\label{fig:bonds}The distributions of the bond densities for the $10^6$
    solutions. Note that these densities are just like random Gaussian
    variables, except for the outliers.}
\end{figure}

The histogram of the density of links for the $10^6$ solutions is
shown in \fref{fig:bonds}. These distributions are obtained for simulations
in which links are flipped randomly in a symmetric fashion. One can
easily push these densities somewhat up or down, by
favoring/restricting the flips of links towards 1. However, much more
extreme solutions can be found by deterministic procedures which turn
as many links to 1 resp.~0. In these cases, we have obtained densities
of as high as 0.96 and as low as 0.14, that is, 2452/2550 links,
resp.~372/2550 links. Two such extreme cases are illustrated in
\fref{fig:fig5b}.
This shows that the model, if needed, can be adapted to questions of
temperature dependence of the protein, for example, by giving more or
less weight to the number of bonds, something like a chemical
potential in statistical mechanics.\\

\begin{acknowledgments} We thank Stanislas~Leibler, Michael R.~Mitchell,
Elisha Moses and Giovanni Zocchi for essential discussions and encouragement. JPE
is supported by an ERC advanced grant `Bridges', and TT by the Institute for Basic Science IBS-R020 and the Simons Center for Systems Biology of the Institute for Advanced Study, Princeton.
\end{acknowledgments}

\bibliography{prx1}
\end{document}